\documentclass[aps,pra,showpacs,twoside,twocolumn,nofootinbib,10pt]{revtex4-2}
\usepackage[colorlinks=true, citecolor=blue, urlcolor=blue ]{hyperref}
\usepackage{epsfig,newlfont,amssymb,amsfonts,amsmath,bm,subfigure,palatino,mathtools,amsthm,braket,soul,enumitem,graphics,graphicx,times,physics, multirow, makecell}
\usepackage{float}

\usepackage[normalem]{ulem}
\usepackage{tabularx}
\usepackage[table,xcdraw]{xcolor}

\begin{document}
\title{
Quantum mutual information as a robust probe of integrability in open quantum systems
}

\author{Nirupam Sen$^{1,2}$, Keshav Das Agarwal$^{1,2}$, Aditi Sen (De)$^{1,2}$}

\affiliation{$^1$Harish-Chandra Research Institute,  Chhatnag Road, Jhunsi, Prayagraj - 211019, India}
\affiliation{$^2$Homi Bhabha National Institute,  Training School Complex, Anushakti Nagar, Mumbai 400 094, India}

\begin{abstract}
 The dynamics of a quantum system encode signatures of whether the underlying Hamiltonian is integrable or chaotic, giving rise to the concept of quantum information scrambling through the properties of the resulting dynamical states or operators.  We introduce an information-theoretic framework based on the Haar-averaged sum of total correlations (aSTC), together with average genuine multipartite entanglement generated dynamically from initially fully separable states, as robust probes of quantum information scrambling. Using the long-range quantum \(XYZ\) spin model in transverse and longitudinal magnetic fields, whose integrable limit is the nearest-neighbor transverse  \(XY\) model, we demonstrate that the long-time average and, more importantly, the temporal fluctuations of the aSTC provide a faithful and system-size-independent signature of integrable and chaotic dynamics, similar to the conventional measure of scrambling, out-of-time-ordered correlator (OTOC). When the system is in contact with the thermal reservoir and system-bath coupling follows Markovianity, we find that the fluctuations of the aSTC and OTOC continue to distinguish integrable and chaotic dynamics only at intermediate times. However, we observe that in the non-Markovian domain, information backflow restores the scrambling dynamics, enabling the aSTC to retain its distinguishing power even at long times. Interestingly, we exhibit that, under Markovian amplitude damping and non-Markovian dephasing noise,  the temporal fluctuations of the aSTC can discriminate between integrability and non-integrability in the weak Markovian regime, even when OTOC fails to do so.
\end{abstract}
\maketitle

\section{Introduction}

The dynamics of interacting quantum many-body systems far from equilibrium lie at the heart of both quantum science and the development of quantum technologies~\cite{Polkovnikov2011_rmp, Luitz2017, LewisSwan2019, Styliaris2021, Harris2022}. A central question in this context is whether the underlying dynamics are integrable or chaotic, since integrability fundamentally governs thermalization, transport, ergodicity, and information propagation in many-body systems~\cite{Deutsch1991, Srednicki1994, Haake2010, DAlessio2016}. Distinguishing integrable from nonintegrable dynamics is therefore important not only for foundational questions, such as the validity of the eigenstate thermalization hypothesis and ergodicity~\cite{Larson2013, DAlessio2016, Rigol2008, Essler2016}, but also for technological applications ranging from quantum simulation and quantum memories to quantum thermal machines~\cite{znidaric_2025, Shtanko2025}. Similarly, quantum chaotic systems possess typical eigenstates~\cite{Bianchi_2022} and can generate Haar random states at long times~\cite{Ghosh_2025_ens}. Consequently, identifying faithful dynamical signatures that discriminate integrable and chaotic quantum evolution remains a fundamental challenge, motivating the present work. 

% Therefore, the distinction between chaotic and integrable systems has lead to eigenspectrum based quantifiers~\cite{Khasseh_2023}. Such quantifiers breaks down in open quantum systems~\cite{Villasenor_2024}.

One of the most distinctive manifestations of many-body dynamics is quantum information scrambling, in which initially localized quantum information disseminates into non-local degrees of freedom through unitary evolution~\cite{Hayden_2007, Sekino_2008, Hosur2016, Maldacena2016, Iyoda2018, Guo2020, Bertini2020, Yan2020, Xu2024_prxq}. Such delocalization provides inherent protection against local errors~\cite{Choi_2020, Rampp2024} and acts as a fundamental quantum resource~\cite{Garcia2023}, since it renders information inaccessible to local observables~\cite{Nahum2018, Keyserlingk_2018}. Further, it is intertwined with quantum chaos, which is traditionally described by the spectral properties of interacting Hamiltonian~\cite{Deutsch1991, Srednicki1994, Haake2010, DAlessio2016} in conjunction with random matrix theory~\cite{Wigner_1951, Mehta1991, Kos_2018}. While black holes~\cite{Hayden_2007, Sekino_2008, Hosur2016, Maldacena2016} and random unitary circuits~\cite{Nahum2018, Bertini2020} are conjectured to be the fastest scramblers in nature, integrable systems can also exhibit scrambling phenomena~\cite{Xu2020, Dowling2023}.
Therefore, scrambling is generally viewed as a necessary, but not strictly sufficient, condition for chaotic (non-integrable) dynamics.

\begin{figure}
  \includegraphics[width=1.0\linewidth]{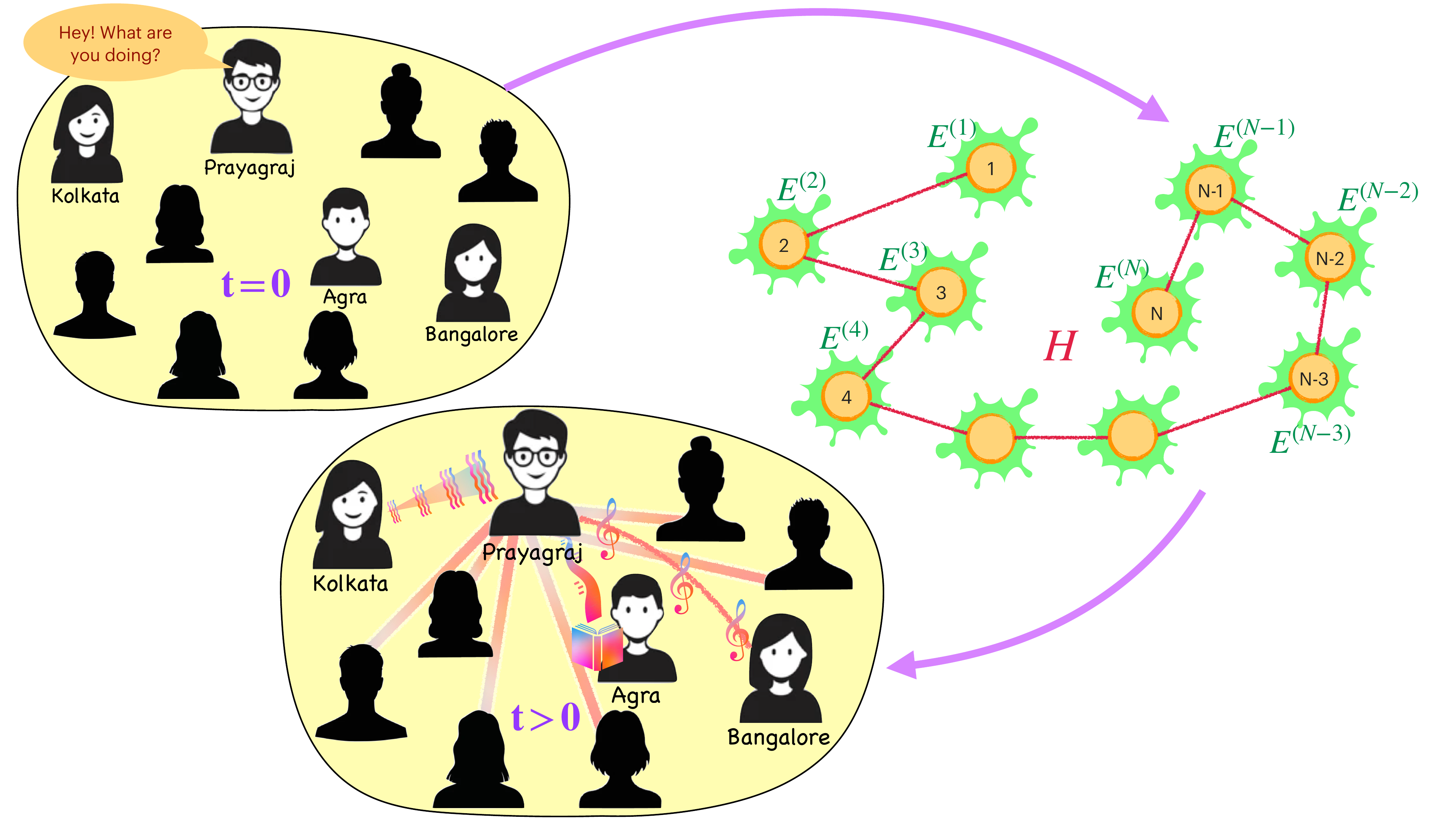}
  \caption{ {\bf Set-up to determine integrability under decoherence}. At initial time \(t=0\), spatially separated parties denoted by  spin-$1/2$ quantum systems at sites $1,2,\dots,N$ are uncorrelated, as shown in the box above.  At a later time, the system interacts through Hamiltonian $H$, thereby creating multipartite quantum correlations among constituent parties due to scrambling. Such quantum information scrambling can be measured through multipartite entanglement or the sum of mutual information between two parties, quantifying total correlations among parties which is illustrated by the box below. Moreover, each site can be attached to a local bath, and hence the evolution can be described by the Gorini-Kossakowski-Lindblad-Sudarshan master equation. The aSTC power of the time-evolved state can be shown to carry the signature of integrability or the chaotic nature of the evolving Hamiltonian. }
\label{fig:schematic}
\end{figure}

Traditionally, quantum information scrambling has been characterized through operator-based quantifiers, most prominently out-of-time-ordered correlators (OTOCs)~\cite{Aleiner2016, Rozenbaum2017, Hashimoto2017, Styliaris2021}, together with operator entanglement~\cite{Prosen_2007b, Bruno2020}, Krylov complexity~\cite{Parker2019, Nandy2025}, and spectral form factors~\cite{Liu_2018, Bertini_2018, Dong_2025}. In particular, the long-time behavior and temporal fluctuations of OTOCs provide a reliable discriminator between integrable and chaotic dynamics~\cite{Liu_2018, Garca_Mata_2018, Fortes_2019, Yoshida_2019, Omanakuttan_2023, Sunil2025, Duarte_2026}. Although experimentally demanding because they require both forward and backward time evolution, OTOCs have nevertheless been measured in several quantum platforms~\cite{Swingle2016, Zhu2016, Landsman2019, Blok2021, Harris2022} and estimated using randomized measurement protocols~\cite{Vermersch_2019, toga2026}. Complementary state-based approaches employing entanglement entropy~\cite{Kim2013, Luitz2017, Nakagawa2018}, Loschmidt echo~\cite{Yan2020}, quantum discord~\cite{Madhok_2015}, quantum coherence~\cite{Anand2021}, and tripartite mutual information with auxiliary~\cite{Rangamani2015, Rota2016} have also been developed. Note, however, that although multipartite correlations naturally emerge as a quantifier for quantum information scrambling and a valuable resource for several quantum technologies, their generation from initially fully separable states as faithful identifiers for distinguishing integrable and chaotic dynamics remains largely unexplored~(cf.~\cite{Pappalardi2018, Bera_2020, Hu_2026}).

In this work, we address this question by proposing information-theoretic diagnostics based on quantum as well as classical correlations generated during the dynamics. Specifically, we investigate the Haar-averaged sum of total correlations (aSTC), defined through the sum of pairwise quantum mutual information~\cite{nielsenchuang, Cerf1997, Groisman2005, Chisholm2024}, together with the Haar-averaged multipartite entangling power~\cite{Mondal_2025, Samanta_2026}, both generated from initially fully separable states evolving under integrable and chaotic many-body Hamiltonians\footnote{Since the initial state is fully separable, the quantum correlation created through the given Hamiltonian can be called the quantum correlation power of the Hamiltonian, and after performing the Haar average, the corresponding quantity can be referred to as the average quantum correlation power~\cite{Zanardi2001,Samanta_2026,Zanardi2000}.}. We demonstrate that these quantities faithfully distinguish integrable and nonintegrable dynamics through their long-time behavior and temporal fluctuations, while naturally quantifying the scrambling of quantum information in closed systems and agreeing with standard OTOC operator metrics. For manifestation, we choose the quantum \(XYZ\) model with variable-range interaction in the presence of transverse and longitudinal magnetic fields.  Furthermore, aSTC power can be experimentally estimated, as identical-copy architectures allow for the direct experimental measurement of quantum purity and mutual information across distant qubits~\cite{Islam2015, Tajik2023, Brydges2019}. 

In realistic settings, quantum systems inevitably interact with their surrounding environment, making it essential to examine the robustness of quantum information scrambling under decoherence. Within the Born--Markov approximation, the reduced dynamics are governed by the Gorini--Kossakowski--Lindblad--Sudarshan (GKLS) master equation~\cite{open_quan_book, Rivas2012, lidar_2020_lecture}, leading to an irreversible loss of information from the system to a memoryless environment. Beyond this approximation, however, finite-size environments, initial system-bath correlations, or strongly time-dependent system-bath interactions~\cite{Chruscinski2010, Rajagopal2010, Benatti2012} give rise to non-Markovian dynamics~\cite{Vacchini_2011, Rivas_2014, Breuer2016_rmp, deVega2017_rmp}, where information can flow back from the environment to the system~\cite{Lu2010, Laine2012, Luo2012, Haseli2014, Fanchini2014}. Such memory effects, quantified by several well-established measures~\cite{Wolf2008, Breuer2009, Rivas2010, Laine2010, Chruscinski2011, Lorenzo2013, Chruscinski2014, Wissmann2015, chanda_2016}, naturally raise the question of whether they can preserve or even enhance the ability of quantum information scrambling to distinguish integrable from chaotic dynamics.

Analyzing the Markovian dynamics, we observe that the irreversible loss of information to a memoryless environment leads to a monotonic decay of both the OTOC and the average sum of total correlations, thereby erasing their ability to distinguish integrable from chaotic dynamics in the long-time limit. Nevertheless, we demonstrate that the temporal fluctuations of the aSTC and OTOC at intermediate times remain sufficient to faithfully distinguish them when the noise is of the dephasing and amplitude damping type. In contrast, non-Markovian dephasing environments qualitatively alter this behavior: the backflow of information from the environment suppresses the loss of correlations, slows the decay of both the OTOC and the aSTC, and restores pronounced long-time temporal fluctuations characteristic of integrable dynamics. Consequently, non-Markovianity, especially non-Markovian dephasing noise, not only preserves quantum information scrambling but also significantly enhances the ability of both aSTC and OTOCs to discriminate between integrable and non-integrable quantum dynamics, which is not pronounced in the case of repetitive system-bath interactions. Furthermore, by investigating quantum information scrambling under Markovian amplitude damping and non-Markovian dephasing noise, we observe that the temporal fluctuations of the aSTC are capable of determining dynamics generated by integrable and non-integrable Hamiltonians in the weak Markovian regime, where the OTOC loses its discriminating power.

The organization of the paper is presented as follows. In Sec.~\ref{sec:infoscrambler}, we introduce the general framework for investigating the integrability of the evolution operator under both unitary and open-system dynamics, together with the quantum information scrambling quantifiers employed in this work. Section~\ref{sec:Markov} presents the behavior of the average sum of total correlations under unitary evolution and subsequently under Markovian dynamics. The effects of non-Markovian dynamics are discussed in Sec.~\ref{sec:non_markov}. We summarize our main findings in Sec.~\ref{sec:conclusion}.

% \begin{enumerate}
%     \item \cite{Luo2012, Haseli2014} - Mutual information based non-Markovianity
%     \item \cite{Lu2010} - Information theoretic quantity for non-Markovianity - QFI.
%     \cite{Islam2015} - Two identical copies enables us to directly measure quantum purity, Rényi entanglement entropy, and mutual information
%     \cite{Tajik2023} - Mutual information experiment across spatially distant qubits
%     \cite{Singh_2026} - Information Storage and Transmission under Markovian Noise - with mutual information
%     \cite{Syzranov_2018, Zhang_2019, Gonzalez_2019, Touil_2021, Zanardi2021, Andreadakis_2023, Bhattacharya2022, Bhattacharjee_2023, Bose2024, Bergamasco_2025, Tripathy_2026} - Markovian scrambling
%     \cite{Han2022, Gribben_2024, Bose_2025} - non-Markovian scrambling
%     \cite{Barch_2023, Ferrari_2025} - non-Hermitian scrambling
%     \cite{Liu_2018, Garca_Mata_2018, Fortes_2019, Yoshida_2019, Omanakuttan_2023, Duarte_2026} - long time dynamics for scrambling 
%     \cite{Touil2020} - Quantum scrambling and the growth of mutual information
% \end{enumerate}

\section{Set-up to study  information scrambling and its quantifiers}
\label{sec:infoscrambler}

We now describe the framework adopted here to determine the integrability of quantum spin models containing $N$ spin-$1/2$ sites. \\
{\it Initialization.} The system is initially prepared in product state as $\rho_0 = \bigotimes_{k=1}^{N} \ket{\psi}_k\bra{\psi}_k$ where $\ket{\psi}_k=c_0\ket{0}_k+c_1\ket{1}_k$ with $c_0$ and $c_1$ being complex numbers, satisfying $|c_0|^2+|c_1|^2=1$. In our analysis, $c_{j}=a_{j}+ia_{j}^\prime$ $(j=0,1)$ are chosen randomly from the Gaussian distribution with vanishing mean and unit standard deviation and followed by normalization, leading to Haar uniform generation of initial pure states. \\
{\it Dynamical state.} For a fixed choice of $(a_j,a_j^\prime)$-pairs $(j=0,1)$ chosen Haar randomly at each site, the initial state $\rho_0$ evolves according to an interacting Hamiltonian, $H$, which may lead to the scrambling of quantum information under unitary evolution. Consequently, the initial product state becomes entangled, and the local quantum information spreads across various degrees of freedom, such that its retrieval generally necessitates access to the complete system~\cite{Luitz2017, Nahum2018, Iyoda2018, Alavirad2019, Xu2024_prxq}. 

In practical scenarios, the system interacts with the environment, thereby leading to non-unitary evolution affecting the information scrambling~\cite{Syzranov_2018, Gonzalez_2019, Singh_2026, Zhang_2019, Zanardi2021, Andreadakis_2023, Han2022}.  We assume that all the sites are connected with local baths, described via Lindblad operators $\{L^{(k)}_{j}\}_{j=1}^{N}$ at each site $j$, with $(k)$ denoting a specific type of environment. Specifically, when the baths remain uncorrelated with the system at the initial time and satisfy the Born-Markov approximations, the system undergoes Markovian dynamics~\cite{open_quan_book, Rivas2012, lidar_2020_lecture} and the evolution of the system can be described by completely positive and trace-preserving (CPTP) operators.
% as the system dynamics is unaffected by the presence of non-interacting auxiliaries, keeping the baths memoryless.
Therefore, the time-evolved state $\rho_t$ can be found by solving the Gorini–Kossakowski–Sudarshan–Lindblad  master equation~\cite{open_quan_book, Rivas2012, lidar_2020_lecture} in the Schr{\"o}dinger picture, 
\begin{align}
\frac{d\rho_t}{dt} &= \mathcal{L}(\rho_t) = -i[H,\rho_t] + \sum_{k,j}\gamma^{(k)}_j\mathcal{D}[L^{(k)}_{j}](\rho_t), \label{eq:gkls_st}\\ 
&\text{with} \quad\mathcal{D}[X](\rho)=X\rho X^{\dagger}
- \frac{1}{2}\{X^{\dagger} X, \rho\}\nonumber,
\end{align}
where $\mathcal{L}(\bm{\cdot})$, and $\mathcal{D}(\bm{\cdot})$ denote the Lindbladian and its dissipative part, respectively while $\gamma^{(k)}_{j}$ is the coupling of the site $j$ of the system with the bath $k$, representing the decay rate. When $\gamma^{(k)}_{j}=0$, the system is isolated, undergoing a unitary evolution, while for the time-dependent $\gamma^{(k)}_{j}$ with $\gamma^{(k)}_{j}<0$ at some time $t$, the system follows non-Markovian dynamics, which will be discussed in Sec.~\ref{sec:non_markov}. 

To illustrate the effect of baths on the operator dynamics, traditionally used to study scrambling via OTOCs, we investigate the dynamics of initially local operators $O_{t=0}$ on the sites of the system, which, in the Heisenberg picture, evolve as
\begin{align}
\frac{dO_t}{dt} =& \mathcal{L}^\dagger(O_t) = i[H,O_t] + \sum_{k,j}\gamma^{(k)}_j\mathcal{D}^{\dagger}[L^{(k)}_{j}](O_t), \label{eq:gkls_op} \\
&\mathcal{D}^{\dagger}[X](O)=X^{\dagger} O X
- \frac{1}{2}\{X^{\dagger} X, O\}\nonumber,
\end{align}
where $\mathcal{L}(\bm{\cdot}),\mathcal{D}(\bm{\cdot})$ and $\gamma^{(k)}_j$ represent the same as described above. To study the scrambling of quantum information by a system, we find $\rho_t$ and $O_t$ by solving Eqs.~(\ref{eq:gkls_st}) and~(\ref{eq:gkls_op}) numerically using Runge-Kutta fourth-order (RK4) method, and study the dynamical properties of states and operators, namely the mutual information in the dynamical states and the OTOCs of operators in the Heisenberg picture. 

The Hamiltonian involved in Eqs.~(\ref{eq:gkls_st}) and~(\ref{eq:gkls_op}) are chosen either integrable or chaotic in nature. In particular, we consider a one-dimensional spin-$1/2$ chain of $N$ sites governed by the (long-range) $XYZ$ interactions,
\begin{align}
H \!=\! \sum\limits_{\substack{i,j=1 \\ i<j}}^{N} &\frac{1}{|i-j|^\alpha}\left[(1\!+\!g)\,\sigma^x_i \sigma^x_j + (1\!-\!g)\,\sigma^y_i \sigma^y_j + \Delta\,\sigma^z_i \sigma^z_j\right]\nonumber \\[-10pt]
     &\quad + h_x \sum_i \sigma^x_i + h_z \sum_i \sigma^z_i, \label{eq:H}
%      \\
% H_{conn} \!=\! &\sum\limits_{i<j} \frac{1}{|i-j|^\alpha}\left[(1\!+\!\gamma)\,\sigma^x_i \sigma^x_j + (1\!-\!\gamma)\,\sigma^y_i \sigma^y_j\right]\prod_{k=i+1}^{j-1}\sigma^z_j\nonumber\\
%      &\quad + h_x \sum_i \sigma^x_i + h_z \sum_i \sigma^z_i,
\end{align}
where $\sigma^\mu_j (\mu\!=\!x,y,z)$ denotes the Pauli matrices acting on site $j$, $g$ and $\Delta$ represent the anisotropy and coupling strength in the $z$-direction, respectively, the interaction strength between spins \(i\) and \(j\) decays polynomially with a fall-off rate $\alpha$, recovering the nearest-neighbor interaction model in the limit $\alpha\gg2$~\cite{Campa2014, Maity2019_review, Defenu2024Jun, lr_rmp_2023}, and $h_x$ and $h_z$ are magnetic field strengths in the $x-$ and $z-$ directions, respectively. The longitudinal magnetic field ($h_x$), $\Delta$, and the long-range interaction strength are critical parameters that govern the onset of chaos~\cite{Shiraishi2019, Shiraishi2024} and the subsequent scrambling of information. Conversely, in the nearest-neighbor limit ($\alpha\to\infty$) with $\Delta=h_x=0$, the Hamiltonian becomes completely integrable and can be exactly solved via the Jordan-Wigner transformation~\cite{lieb1961, barouch_pra_1970_1, barouch_pra_1970_2, glen2020} for a given anisotropy ($g$). Let us denote the parameter set $ p_1 =\{g\!=\!0.5, \ \alpha\!=\!10, \ h_z\!=\!1, \ \Delta\!=\!h_x\!=\!0\}$, where the model is integrable. Note that $\alpha=10$ typically mimics the system with the nearest-neighbor interactions only. To understand the effects of different parameters on information scrambling, we study the individual roles of $\Delta\neq0, h_x\neq0$ and LR interactions with $\alpha <2$. For example, we consider the non-integrability of the model when $h_x=1$ and all other parameters are given by $p_1$.

% To characterize the integrability and chaotic nature of system via scrambling dynamics, we study the total correlations generated in the quantum system, quantified by the quantum mutual information between the sites of the system.

After obtaining  $\rho_t$ for a fixed Haar-randomly chosen initial state (which we refer to as a single realization), our goal is to compute multi-site quantum correlation quantifiers, $\mathcal{Q}(\rho_t)$. 
%that are capable of distinguishing chaotic systems from integrable ones. Finally, 
The dependence of the initial state on $\mathcal{Q}(\rho_t)$ can be removed by performing a Haar average on the initial product state, i.e., we calculate
\begin{equation}
     \langle\mathcal{Q}\rangle \equiv \langle\mathcal{Q}(\rho_t)\rangle = \int_{ \ket{\psi}\in \substack{\text{Haar}\\\text{sample}}} d\left(\mathop{\otimes}_{k=1}^{N}\ket{\psi}_k\right)\;\mathcal{Q}(\rho_t),
     \label{eq:HaaravQ}
\end{equation}
where the integration is performed over all initial pure states $\ket{\psi}_k$ sampled Haar randomly. For our analysis, depending on the system-size, we choose $100$ to $1000$ realizations at each site to obtain $\langle\mathcal{Q}\rangle$. Our aim is to identify a Haar-averaged multipartite quantum correlation quantifier that is capable of distinguishing dynamics generated through chaotic systems from integrable ones. 

In this work, we mainly focus on the strength of a variant of quantum mutual information~\cite{Groisman2005} and a geometric measure of multipartite entanglement~\cite{Wei2003,aditi_pra_2010} for detecting integrable and chaotic systems. Note that $\langle\mathcal{Q}\rangle$ represents the average quantum correlation power~\cite{Zanardi2000,Zanardi2001,Samanta_2026} which, typically, characterizes the features of an operator, in this case, the Hamiltonian, under consideration.

\subsection{Average sum of quantum mutual information: An identifier for integrability}

Quantum mutual information quantifies the correlation constituting both classical and quantum correlation of a bipartite quantum state $\rho$ shared between two parties $1$ and $2$~\cite{nielsenchuang, Cerf1997, Groisman2005}. It is defined as $I_{\rho}(1:2) = S(\rho_1) + S(\rho_2) - S(\rho)$, where $S(\nu)=-\text{Tr}\left(\nu \log_2\nu\right)$ denotes the von Neumann entropy of the state $\nu$ and $\rho_x(x=1,2)$ are the reduced density matrices of the bipartite state $\rho$. In a spin chain of $N$ sites, to quantify how the correlations spread over time, we compute a quantity, the sum of bipartite total correlations (STC) shared by a fixed site, say $1$, with the remaining sites. Mathematically,
\begin{align}
\mathcal{I}^{(s)}_{\rho}= \sum_{j=2}^{N} I_{\rho_{1,j}}(1:j), 
\end{align}
where $I_{\rho_{1,j}}(1:j)$ represents the quantum mutual information of the reduced state $\rho_{1,j}$ between the sites $1$ and $j$, obtained from the evolved state $\rho_t$ of $N$ sites. Therefore, the $\mathcal{I}^{(s)}_{\rho}$, quantifies the spread of quantum information from one end of the system to various individual degrees of freedom, generated via the given evolving Hamiltonian $H$, from the initial separable states, thereby carrying the signature of the given Hamiltonian or the evolution operator. 

% \kda{By leveraging the quantum information-theoretic properties of the dynamical states, this metric serves as a natural, intrinsic probe of the system that requires no auxiliary coupling. -- {\it have to discuss separately here and in intro. }}

The information spreading in quantum systems by evolution can depend on the initial configuration of the state $\rho_0$, which can result in the false signature of integrable dynamics for chaotic systems~\cite{Iyoda2018, Xu2020, Dowling2023}. Therefore, to remove the initial state-dependence, we compute the Haar average of STC $\langle\mathcal{I}^{(s)}_{\rho}\rangle$, over initially pure product states (see also Eq.~(\ref{eq:HaaravQ})) to faithfully quantify the state-independent scrambling of information by system $H$. A key advantage of quantum mutual information is that it can be readily evaluated in open quantum systems, where interactions with the environment drive the system into mixed states over time.

%Specifically, the initial state of the system $\rho_0 = \bigotimes_{k=1}^{N} \ket{\psi_k}\bra{\psi_k}$,
% with $\ket{\psi_k}=c_0\ket{0}_k+c_1\ket{1}_k$ are chosen Haar randomly over the Bloch sphere, i.e., $c_0$ and $c_1$ are sampled from Gaussian distribution with $|c_0|^2+|c_1|^2=1$. For our study, we perform the average over $1000, 800, 500, 300, 250, 200$ and $100$ Haar sampled pure product states for $N=6,8,10,12,14,16$ and $18$ system-sizes respectively and compute ensemble averaged STC
% \begin{equation}
%     \langle\mathcal{I}^{(s)}\rangle(t) = \int_{\text{Haar}}\!\!\!\!\!d\ket{\psi}_k \;\mathcal{I}^{(s)}_{\rho_t}\bigg|_{\rho_0=\bigotimes_{k=1}^{N} |\psi\rangle_k\bra{\psi}},
% \end{equation}
% where $\rho_t$ is the dynamical state evolving according to Eq.~(\ref{eq:gkls_st}). 
To probe the role of multipartite quantum correlations in determining the integrability of the system, we also consider
 the Haar-average genuine multipartite entanglement (GME) based generalized geometric measure \cite{GGM_wei2003, aditi_pra_2010, Samanta_2026} under unitary dynamics. It can  be shown to distinguish the dynamics generated via the integrable systems from the chaotic ones (see Appendix~\ref{app:ggm}). However, extending such GME-based characterization to noisy environments possesses a significant challenge, as evaluating the GME for mixed states is computationally intractable. Consequently, we will demonstrate that the  aSTC  serves as a viable alternative for determining the properties of the Hamiltonian.

\subsection{Out-of-time-ordered correlators: Conventional measure of information scrambling}

To verify whether quantum information theoretic quantities can capture the integrability of the system, especially in the presence of environmental interactions with the system, we also calculate the traditional quantifier, out-of-time-ordered correlators, which quantifies the spread of local operators during the dynamics by the rate of spread of the support of the local operators~\cite{Parker2019, Bertini2020}. For two local operators $V_0$ at the initial time and $W$ on spatially separated sites, OTOC is defined as the support of the time-evolved operator $V_t$ on the support of $W$, given by
\begin{equation}
\mathcal{C}(t)=\frac{1}{2D}\text{Tr}([V_t,W]^\dagger[V_t,W]),
\end{equation}
where $D$ is the dimension  of Hilbert space. The OTOC provides an upper bound over the information of $V_0$ propagated from a fixed site and measured by another site with $W$, which are typically chosen to be far from the given site on which $V_t$ is measured~\cite{Xu2024_prxq}. Therefore, while it gives a state-independent quantifier of information scrambling, it is an operator-dependent quantifier for the choice of $V_0$ and $W$. To validate the role of average STC, we compute OTOC with (arbitrarily chosen) $V_0=\sigma_{2}^x $ and $W=\sigma_{N-1}^y$, both of  which are situated at the second sites from the ends of the open chain. Since the operators $V_0$ and $W$ act on different local Hilbert spaces, they commute initially (at time $t=0$), but as time progresses, the support of $W_t$ spreads across the system, giving a non-vanishing value of OTOC.

\begin{figure}
  \includegraphics[width=\linewidth]{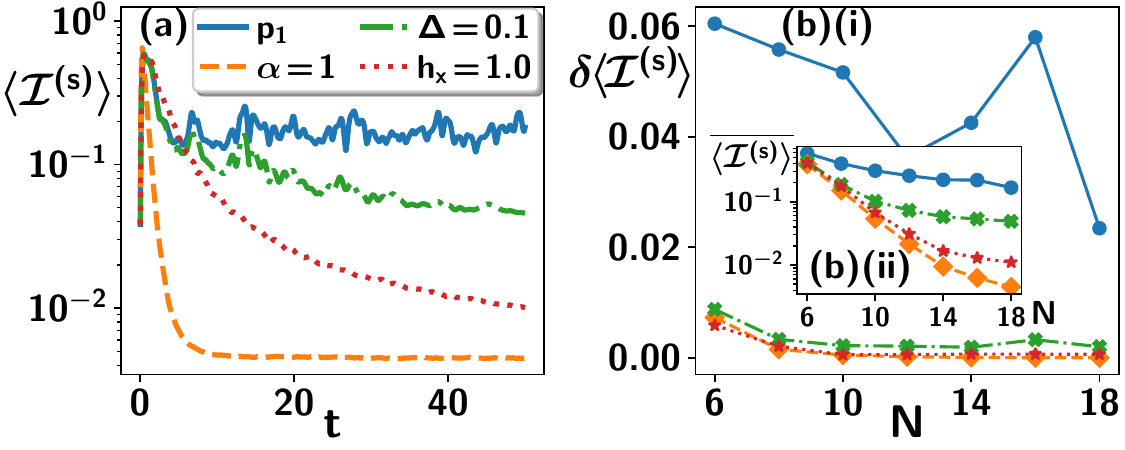}
  \caption{{\bf Dynamics of average STC power, $\langle\mathcal{I}^{(s)}\rangle$ for integrable and chaotic systems in a noiseless scenario.} (a) $\langle\mathcal{I}^{(s)}\rangle$ (ordinate in log-scale) against time (abscissa) for unitary evolution with Hamiltonian $H$ in Eq.~(\ref{eq:H}), with various parameters represented by different line styles (and color) for system-size $N=18$. The integrable system (with parameter $p_1$) is given by solid (blue) line, whereas other parameters are mentioned only with the value which is different from $p_1$.  The temporal fluctuations $\delta\langle\mathcal{I}^{(s)}\rangle$ are high for $p_1$, and are vanishing for chaotic systems at long times. (b) The dependence of $\delta\langle\mathcal{I}^{(s)}\rangle$ (ordinate) on the system-size (abscissa)  (with $(t_0, t_1)=(30, 50)$). The fluctuations $\delta\langle\mathcal{I}^{(s)}\rangle\sim 10^{-2}$ for integrable systems, whereas $\delta\langle\mathcal{I}^{(s)}\rangle\sim 10^{-3}$ for chaotic systems. (Inset) The mean averaged STC power $\overline{\langle\mathcal{I}^{(s)}\rangle}$ (ordinate) with system-size $N$ (abscissa). All the axes are dimensionless.}
  \label{fig:smi_uni}
\end{figure}
\begin{figure}
  \includegraphics[width=\linewidth]{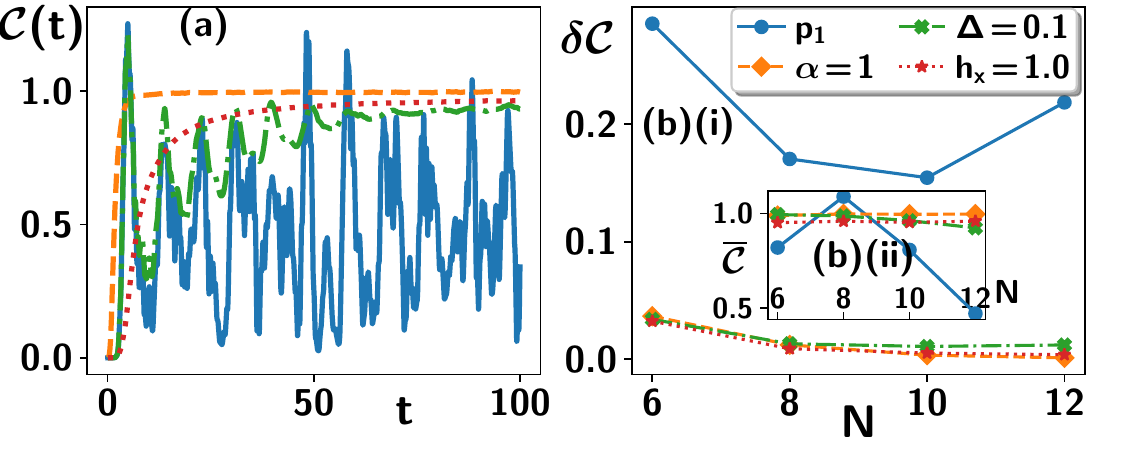}
  \caption{{\bf Dynamics of OTOC under unitary evolution}. (a) $ \mathcal{C}(t)$ (ordinate) vs time (abscissa) for  $N=12$. Different lines correspond to the temporal behavior of OTOC for Hamiltonian $H$ in Eq.~(\ref{eq:H}), with various parameters. The integrable system  with \(p_1\) (solid line) produces a high temporal fluctuation, whereas others show suppression of fluctuation in the long time. Behavior is very similar with aSTC  $\langle\mathcal{I}^{(s)}\rangle$ in Fig.~\ref{fig:smi_uni}(a). (b) $\delta \mathcal{C} $ (ordinate)  against $N$ (abscissa)  for  $(t_0, t_1)=(60, 100)$. The fluctuation $\delta \mathcal{C} \sim O(10^{-1}) $ for the integrable system ($p_1$), while $\delta \mathcal{C} \sim O(10^{-2})$ for all chaotic cases. Except $p_1$, all chaotic systems show vanishing $\delta \mathcal{C}$ with increasing \(N\), the same as depicted in Fig.~\ref{fig:smi_uni}(b). (Inset) The mean of \(\mathcal{C}\), $\bar{\mathcal{C}}$ (ordinate) with system-size $N$ (abscissa), showing temporal fluctuation in $p_1$ is independent of system-size. All axes are dimensionless.}
  \label{fig:otoc_uni}
\end{figure}

\begin{figure}
  \includegraphics[width=\linewidth]{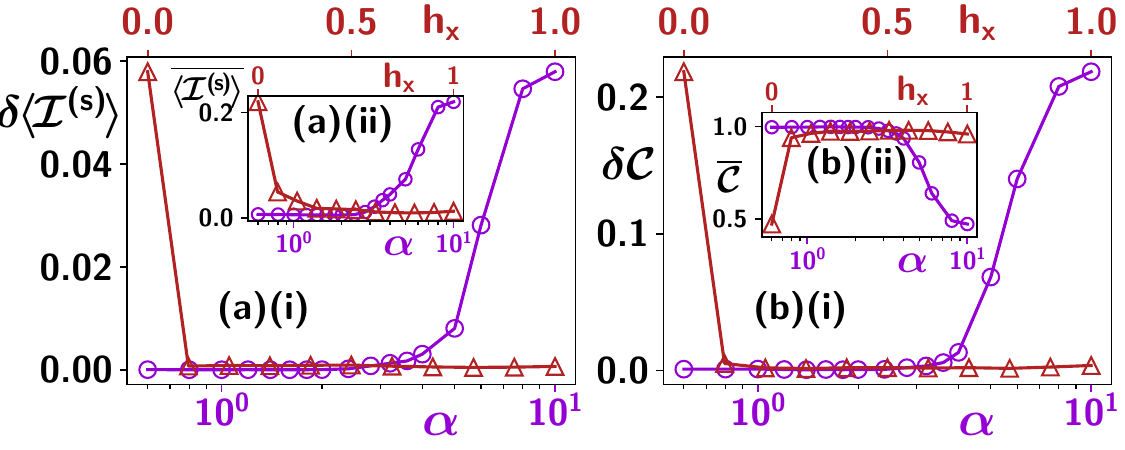}
  \caption{{\bf Dependence  of aSTC and OTOC on  $\alpha$ and $h_x$ in the unitary dynamics  }. (a) For $N=16$, $\delta\langle\mathcal{I}^{(s)}\rangle$ (ordinate) against $\alpha$ (lower abscissa) and $h_x$ (upper abscissa). The integration is performed from $t_0=30$ to $t_1=50$. As $\alpha$ increases, thereby approaching the integrable systems, $\delta\langle\mathcal{I}^{(s)}\rangle$ also increases, while increasing $h_x$ shows completely opposite behavior. (Inset) The mean aSTC $\overline{\langle\mathcal{I}^{(s)}\rangle}$ (ordinate) with $\alpha$ and $h_x$ (abscissa). (b) $\delta \mathcal{C}$ (ordinate) with $\alpha$ and $h_x$ (abscissas) for $(t_0, t_1)=(60, 100)$. $\delta \mathcal{C}$ also shows very similar feature like $\delta\langle\mathcal{I}^{(s)}\rangle$, it increases with long-range interaction strength $\alpha$ and decreases, when  $h_x$ increases. (Inset) The mean OTOC $\bar{\mathcal{C}} $ (ordinate) with $\alpha$ and $h_x$ (abscissas). Here \(N=12\). All axes are dimensionless.}
  \label{fig:smi_otoc_uni}
\end{figure}

\section{Integrability under Markovian dynamics}
\label{sec:Markov}

In the presence of environmental interactions, the scrambling of quantum information is expected to be suppressed~\cite{Syzranov_2018, Gonzalez_2019, Zhang_2019, Zanardi2021} due to the flow of information from the system to the environment. In this section, we illustrate how Markovian baths can influence the dynamics of the average STC power, which can, in this way, highlight its efficacy in distinguishing the dynamics of the integrable system from the chaotic one. Before presenting the results for Markovian baths, let us first concentrate on the noiseless scenario, i.e., when the system undergoes a unitary evolution, which actually can be used as a benchmark for the noisy scenarios.

\subsection{Noiseless scenario: Capability of average STC power for integrability }

Let us assume that the system is isolated which implies $\gamma^{(k)}_j=0 \ \forall j$ in Eqs.~(\ref{eq:gkls_st}) and~(\ref{eq:gkls_op}) and hence the evolution is unitary. In this limit, the question of integrability has been extensively explored using out-of-time-ordered correlators~\cite{Luitz2017, Hashimoto2017, Iyoda2018, Alavirad2019, Xu2024_prxq}. Previous studies have established that OTOCs can distinguish between integrable and chaotic models through fluctuation of operator dynamics at long times~\cite  {Liu_2018, Garca_Mata_2018, Fortes_2019, Yoshida_2019, Omanakuttan_2023, Sunil2025, Duarte_2026}. 

We demonstrate that the Haar-averaged STC, $\langle\mathcal{I}^{(s)}\rangle$ and average multipartite entangling power $\langle\mathcal{G}\rangle$ (see Appendix~\ref{app:ggm}) can also exhibit characteristic dynamics that clearly differentiate evolution governed by an integrable Hamiltonian from chaotic ones. Since the unitary dynamics preserves the purity of an initial pure state, we can simulate larger system sizes up to $N=18$ (see Appendix~\ref{app:chb}), by varying the parameters of the Hamiltonian $H$ in Eq.~(\ref{eq:H}) from the integrable limit $p_1 =\{\gamma=0.5, \alpha=10, h_z=1, \Delta=h_x=0\}$ to non-integrable ones. Note that while mentioning the parameters, we indicate only those parameter values, which differ from the set of values included in $p_1$.

Under unitary dynamics initiated from Haar-random pure separable states, mutual information spreads across the system, resulting in $\langle\mathcal{I}^{(s)}\rangle$ to exhibit rapid growth at early times. At large times, it oscillates with time around a non-vanishing value for the integrable Hamiltonian, while it decays for chaotic system without any oscillations (see Fig.~\ref{fig:smi_uni}(a)). Similar behaviors emerges for average multipartite entanglement power $\langle\mathcal{G}\rangle$ (see Appendix~\ref{app:ggm}) and for the OTOC, $\mathcal{C}(t)$, as shown in Fig.~\ref{fig:otoc_uni}(a), where parameter regimes producing fluctuations in the average STC power display corresponding fluctuations in the OTOC. 
% Specifically, in an integrable system, the Haar-averaged STC exhibits large fluctuations, much like the OTOC, whereas in chaotic systems that exhibit information scrambling, $\langle\mathcal{I}^{(s)}\rangle(t)$ decays and $\mathcal{C}(t)$ saturate with minimal fluctuations. 
To quantify this large time fluctuating behavior for a given observable $\mathcal{Q}(t)$, we define first and second moments of $\mathcal{Q}(t)$, as
\begin{align}
    \overline{\mathcal{Q}} = \frac{1}{t_1-t_0}\int_{t_0}^{t_1} \mathcal{Q}(t)dt, \quad \delta \mathcal{Q} = \sqrt{\overline{\mathcal{Q}^2}-(\overline{\mathcal{Q}})^2},
    \label{eq:moments}
\end{align}
respectively. Here, $t_0$ is chosen to be sufficiently large to bypass transient effects and capture the steady-state domain, with $t_1 > t_0$. Ideally, we are interested in the limit $t_1 \to \infty$, but finite $t_1$ is chosen for practical scenarios.
%; specifically, \sout{$t_0=30$ and $t_1=50$ are taken throughout this work, unless stated otherwise.}

We observe that the behavior of both the average STC and its fluctuations in the long-time limit of unitary dynamics can determine the Hamiltonian's properties. Specifically, the fluctuations $\delta\langle\mathcal{I}^{(s)}\rangle$ show a clear distinction between the integrable ($\delta\langle\mathcal{I}^{(s)}\rangle\sim O(10^{-2})$) and chaotic ($\delta\langle\mathcal{I}^{(s)}\rangle\lesssim10^{-3}$) systems for all system-sizes.
However, the saturating value $\overline{\langle\mathcal{I}^{(s)}\rangle}$ of aSTC decreases with increasing system-size due to finite $t_1$, while the decrease is slower for an integrable system than for the chaotic case (see Fig.~\ref{fig:smi_uni}(b)). Therefore, this suggests that $\overline{\langle\mathcal{I}^{(s)}\rangle}$ separates the integrable dynamics from the chaotic ones only at large system-sizes. This result is also in agreement with OTOC, where $\overline{\mathcal{C}}$ is system-size dependent and fluctuations $\delta\mathcal{C}$ of the OTOC is system-size independent, which are large in the integrable dynamics ($\delta\mathcal{C}\sim 10^{-1}$) while near-zero fluctuations ($\delta\mathcal{C}\lesssim 10^{-2}$) is observed in the chaotic case (see Fig.~\ref{fig:otoc_uni}(b)). 
Note that the chaotic features in the Hamiltonian are incorporated either by introducing a longitudinal field, or by $\Delta\neq0$ or by long-range  (LR) interactions, i.e., $\alpha\ll 10$. 

Let us examine how chaotic dynamics are driven only by the LR interaction strength, $\alpha$, and the longitudinal field, $h_x$. If we concentrate on the situation when $h_x=0$, i.e., chaotic feature emerges only due to LR interaction, $\alpha$, 
% As depicted in Fig.~\ref{fig:smi_otoc_uni}, 
the fluctuations, $\delta\langle\mathcal{I}^{(s)}\rangle$, and $\delta\mathcal{C}$ remain almost vanishing for $\alpha \lesssim 2$ but increase continuously for $\alpha > 2$.
Similarly,  introducing a non-vanishing longitudinal field ($h_x > 0$) in \(p_1\) strictly suppresses both $\delta\langle\mathcal{I}^{(s)}\rangle$ and $\delta\mathcal{C}$ to near zero, signaling a transition to chaos in the Hamiltonian $H$. While the saturation values $\overline{\langle\mathcal{I}^{(s)}\rangle}$ and $\overline{\mathcal{C}}$ also reflect distinct behavior for integrable and chaotic dynamics, they behave differently under varying field strengths. Specifically, $\overline{\langle\mathcal{I}^{(s)}\rangle}$ decreases continuously as $h_x$ increases, whereas the fluctuations remain near zero (see Fig.~\ref{fig:smi_otoc_uni}). Consequently, the fluctuation $\delta\langle\mathcal{I}^{(s)}\rangle$ provides a more reliable indicator of integrability-breaking than the mean value $\overline{\langle\mathcal{I}^{(s)}\rangle}$, which exhibits significant dependence on both system-size and underlying parameters.

\begin{figure}\includegraphics[width=\linewidth]{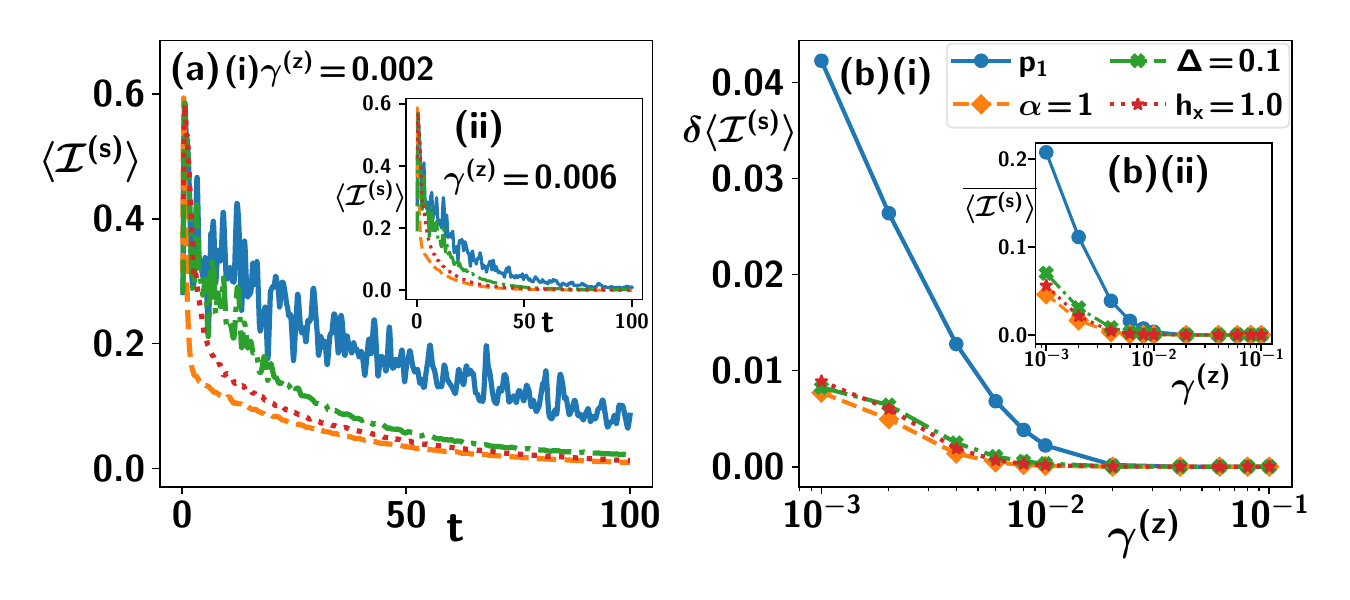}
\caption{ { \bf Dynamics of average STC in the presence of dephasing Markovian noise.}  (a) $\langle\mathcal{I}^{(s)}\rangle$ (ordinate) against time (abscissa) for Markovian dynamics with $\gamma^{(z)}=0.002$ and $N=8$. All other parameters are same as in Fig. \ref{fig:smi_uni}. %Different line-styles (and color) denote the systems with Hamiltonian in Eq.~\ref{eq:H} with various parameters. 
Unlike unitary evolution,  $\langle\mathcal{I}^{(s)}\rangle$ grows initially, but as time progresses further, it starts to decay, although temporal fluctuation in the integrable system $p_1$ ((blue) solid line) still exists, thereby discriminating integrable systems from nonintegrable one in intermediate time. (Inset) $\langle\mathcal{I}^{(s)}\rangle$(ordinate) against  time (abscissa) for $\gamma^{(z)}=0.006$. %$\langle\mathcal{I}^{(s)}\rangle$ decays fast as $\gamma^{(z)}$ increases. 
(b)  $\delta\langle\mathcal{I}^{(s)}\rangle$ and (inset) $\overline{\langle\mathcal{I}^{(s)}\rangle }$ (ordinate) vs decay rate $\gamma^{(z)}$ (abscissa), with $(t_0, t_1)=(60, 100)$. For small $\gamma^{(z)}$ ($\leq 10^{-2}$), integrable system ($p_1$) produces high $\delta \langle\mathcal{I}^{(s)}\rangle$ ($\delta\langle\mathcal{I}^{(s)}\rangle \sim O(10^{-2})$) compared to chaotic cases ($\delta\langle\mathcal{I}^{(s)}\rangle \sim O(10^{-3})$) but at high $\gamma^{(z)} $ values,  $\delta \langle\mathcal{I}^{(s)}\rangle$ vanishes. Furthermore, at small $\gamma^{(z)}$($\leq10^{-2}$), for integrable systems, $\overline{\langle\mathcal{I}^{(s)}\rangle }$ $\sim O(10^{-1})$ while for chaotic cases,  $\overline{\langle\mathcal{I}^{(s)}\rangle }$ $\sim O(10^{-2})$. Similar results can also be obtained for local amplitude damping noise (see Fig. \ref{fig:smi_amp_damp}). All axes are dimensionless.}
  \label{fig:smi_deph}
\end{figure}

\begin{figure}\includegraphics[width=\linewidth]{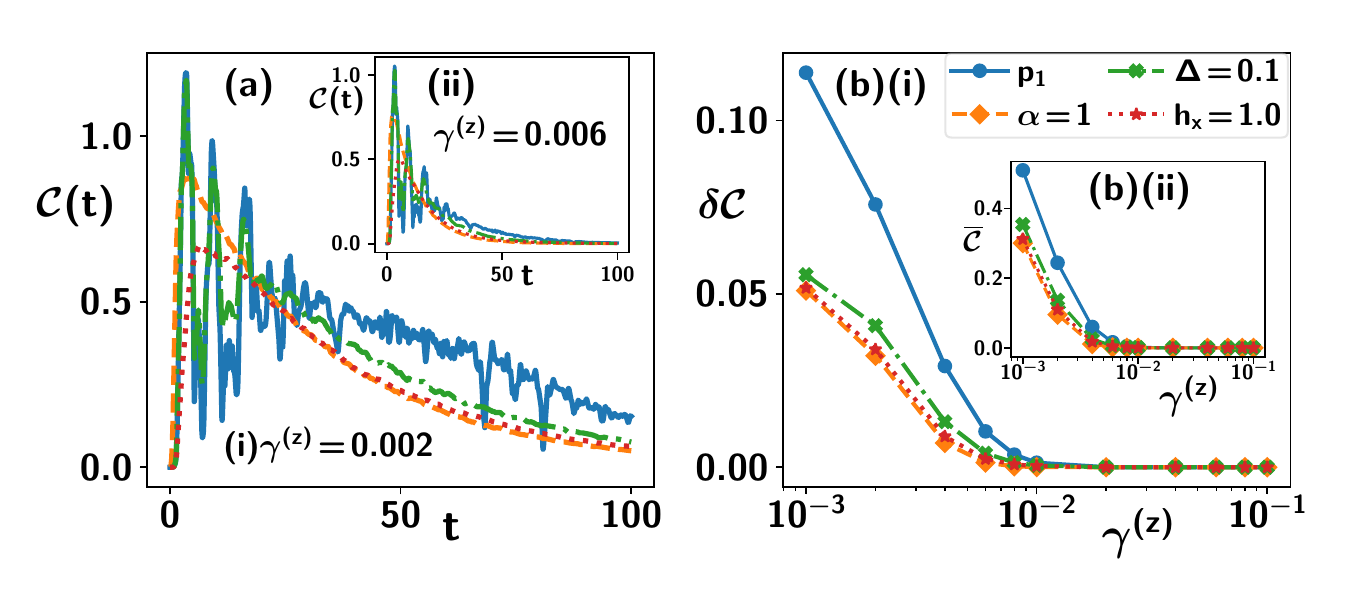}
  \caption{ {\bf OTOC in the presence of dephasing Markovian noise}.  (a) $\mathcal{C}(t)$ (ordinate) with respect to \(t\) (abscissa) for Markovian dynamics with $\gamma^{(z)}=0.002$ and \(N=8\). All other features are the same as in Fig. \ref{fig:smi_deph}.
  %Initially, $C(t)$ grows for all cases, but at long time they all starts to decay. Not only the integrable system ($p_1$) but also the chaotic system with $\Delta=0.1$ initially produces temporal fluctuations. 
  (Inset) For a large value $\gamma^{(z)}=0.006$, $\mathcal{C}(t)$ (ordinate)  decays faster with time, and its capability in determining integrability is lost, as also  shown in Fig. \ref{fig:smi_deph}. 
  %$C(t)$ decays fast and temporal fluctuation of $p_1$ and $\Delta=0.1$ are reduced in short time. 
  (b) $\delta \mathcal{C}$  and \(\overline{\mathcal{C}}\) (inset)(ordinate) vs $\gamma^{(z)}$ (abscissa) (with $(t_0, t_1)=(60, 100)$).  A similar feature emerges as depicted in Fig. \ref{fig:smi_deph}.  See also Fig. \ref{fig:smi_amp_damp} for AD Markovian noise. 
  % At small $\gamma^{(z)}$($\leq 10^{-2}$), for all integrable or chaotic systems  $\delta C\sim O(10^{-2})$ and as $\gamma^{(z)}$ increases more $\delta C$ vanishes for all cases. (Inset) The mean $\overline{C}$ (ordinate) against $\gamma^{(z)}$. With increasing $\gamma^{(z)} $ it starts to decay and vanishes. 
  %The distinction between integrable and chaotic systems (which was present in unitary dynamics) has been lost.
  All axes are dimensionless. }
  \label{fig:otoc_deph}
\end{figure}

\subsection{Markovian dephasing bath: Eradicating integrability-determining feature}
\label{subsec:Markoviandph}

Realistic quantum systems inevitably interact with their surroundings, fundamentally altering their dynamics. Assuming an uncorrelated and memoryless environment, an open-system evolution is governed by Markovian dynamics, described by Eqs.~(\ref{eq:gkls_st}) and~(\ref{eq:gkls_op}) for the states and operators respectively, where specific environmental interactions are captured by distinct Lindblad operators, $L^{(k)}$. For instance, a local dephasing bath at site $j$ is characterized by the Pauli operator $L^{(z)}_{j}=\sigma_{j}^{z}$ with dephasing strength $\gamma^{(z)}$. Conversely, amplitude damping noise at the $j$-th qubit is modeled by the non-Pauli operator $L^{(\pm)}_{j}=\sigma_{j}^{\pm}$, with dissipation coefficient $\gamma^{(\pm)}$ and $\sigma^{\pm}_j = (\sigma^{x}_j \pm i\sigma^{y}_j)/2$. Note that these two baths belong to fundamentally different classes of operators, as dephasing Lindblad operators are described by Pauli operators whereas amplitude damping ones are non-Pauli operators~\cite{Preskill1998, lidar_2020_lecture}. Here, we utilize STC to investigate the effects of dephasing on quantum information scrambling (see Appendix~\ref{app:amp_damp} for the effects of an amplitude damping environment).

% In amplitude damping, the system loses or gains energy due to the interaction with the environment, whereas in dephasing, the system retains its energy, but random phases are added to the coherent components of the state.

In the presence of Markovian baths, the environment dissipates the quantum correlations generated by unitary dynamics, which becomes pronounced for large time. Consequently, while the Haar-averaged STC, $\langle\mathcal{I}^{(s)}\rangle$, initially grows during the transient regime, it ultimately decays at long times (see Fig.~\ref{fig:smi_deph}(a)). This decay reflects a fundamental competition between internal system interactions and environmental dissipation~\cite{open_quan_book}, where the baths suppress the mutual information created among the initial product states. As the system-bath coupling $\gamma^{(z)}$ increases, the STC decays more rapidly, suppressing both its late-time mean value and its temporal fluctuations. Nevertheless, in the weak-coupling regime, the STC remains a viable probe for distinguishing integrable from chaotic dynamics. For instance, at $N=8$ and $\gamma^{(z)}\lesssim 10^{-2}$ (with $t_0=60$ and $t_1=100$), the integrable model (parameter set $p_1$) exhibits fluctuations $\delta\langle\mathcal{I}^{(s)}\rangle\sim O(10^{-2})$ and a mean $\overline{\langle\mathcal{I}^{(s)}\rangle}\sim O(10^{-1})$. In stark contrast, the chaotic Hamiltonian yields significantly lower values of $\delta\langle\mathcal{I}^{(s)}\rangle\sim O(10^{-3})$ and $\overline{\langle\mathcal{I}^{(s)}\rangle}\sim O(10^{-2})$ (see Fig.~\ref{fig:smi_deph}(b)). Expanding the observation window $t_1$ lowers the critical coupling threshold $\gamma^{(z)}$ beyond which this distinction vanishes. Ultimately, strong system-bath couplings blur the dynamical signatures of integrability~\cite{Zanardi2021}, posing a significant challenge for detecting quantum information scrambling in practical experimental setups~\cite{Swingle2016, Zhang_2019, Gonzalez_2019}.

Recent studies have investigated the behavior of OTOCs in the presence of Markovian baths~\cite{Zhang_2019, Gonzalez_2019, Zanardi2021}, revealing that the OTOC decays over time for both integrable and chaotic models, with the faster decay as the system-bath coupling increases. We also find that Markovian dissipation quickly washes out the differences between integrable and non-integrable models while isolated unitary dynamics yield distinct temporal signatures, thereby distinguishing them. For instance, in an integrable model $H$ (Eq.~(\ref{eq:H}), parameter set $p_1$ for $N=8$) subjected to a weak coupling of $\gamma^{(z)}=0.002$, the fluctuations are suppressed to $\delta \mathcal{C}\sim O(10^{-2})$, down from $\delta\mathcal{C}\sim 0.2$ in the unitary case. Conversely, the non-integrable extensions exhibit $\delta\mathcal{C}\sim O(10^{-2})$ regardless of whether the system is isolated ($\gamma^{(z)}=0$) or weakly coupled ($\gamma^{(z)}=0.002$). As the bath coupling is further increased, $\delta\mathcal{C}$ decays continuously, ultimately vanishing for $\gamma^{(z)}\gtrsim 10^{-2}$. The time-averaged value, $\bar{\mathcal{C}}$, similarly decays with increasing coupling strength across all models, failing to distinguish the underlying integrability of the Hamiltonian. We further note that $\bar{\mathcal{C}}\sim 0$, which mathematically necessitates $\delta\mathcal{C}=0$, signifying a complete loss of correlation between the local operators $\sigma_2^x(t)$ and $\sigma_{N-1}^y$.

\section{Benefit of non-Markovian dynamics in determining integrability and non-integrability}
\label{sec:non_markov}

Let us relax the assumption of the Markovian environments ~\cite{Vacchini_2011, Rivas_2014, Breuer2016_rmp, deVega2017_rmp, Xu2026_rmp}, resulting in information backflow from the baths to the system~\cite{Lu2010, Laine2012, Luo2012, Haseli2014, Fanchini2014}. In the presence of non-Markovian baths, various quantum resources, including entanglement~\cite{Rivas2010} and quantum discord~\cite{Fanchini2010}, are shown to revive with time after collapse~\cite{Lu2010, Laine2012, Luo2012, Haseli2014, Fanchini2014}, thereby rising a possibility to mitigate noise. Indeed, it has been reported that several quantum devices can retain their quantum advantages in the presence of non-Markovian noise which typically vanish with Markovian noise~\cite{Srijon2021, sen2023noisybattery}. Therefore, we can also expect that the decaying behavior observed with Markovian baths which eliminate the distinguishing capability of aSTC and OTOC can be recovered when the system-bath interactions are non-Markovian. For example, recently, it has been found that the reflux of information to the system is shown to alter the information scrambling in non-Markovian systems~\cite{Han2022, Gribben_2024, Bose_2025}. We address the role of non-Markovianity in the context of integrability and chaotic systems in four distinct ways -- (1) time-dependent system-bath interactions, i.e., $\gamma_j^{(k)}$ are taken to be time-dependent; (2) non-Markovian dephasing bath along with Markovian amplitude damping bath; (3) repetitive interactions, where some of the sites interact with local finite thermal baths for a small time, say $\tau$, which can be changed to drive the system from Markovian to non-Markovian regimes; (4) the initial product state $\ket{\Psi(0)}=\bigotimes_{k=1}^N\ket{\psi_k}$ influenced by Markovian and non-Markovian noises.
% In this section, we examine the role of non-Markovian dynamics in quantum scrambling from the behavior of STC in the systems, as well as from the dynamics of OTOC.

\begin{figure}\includegraphics[width=\linewidth]{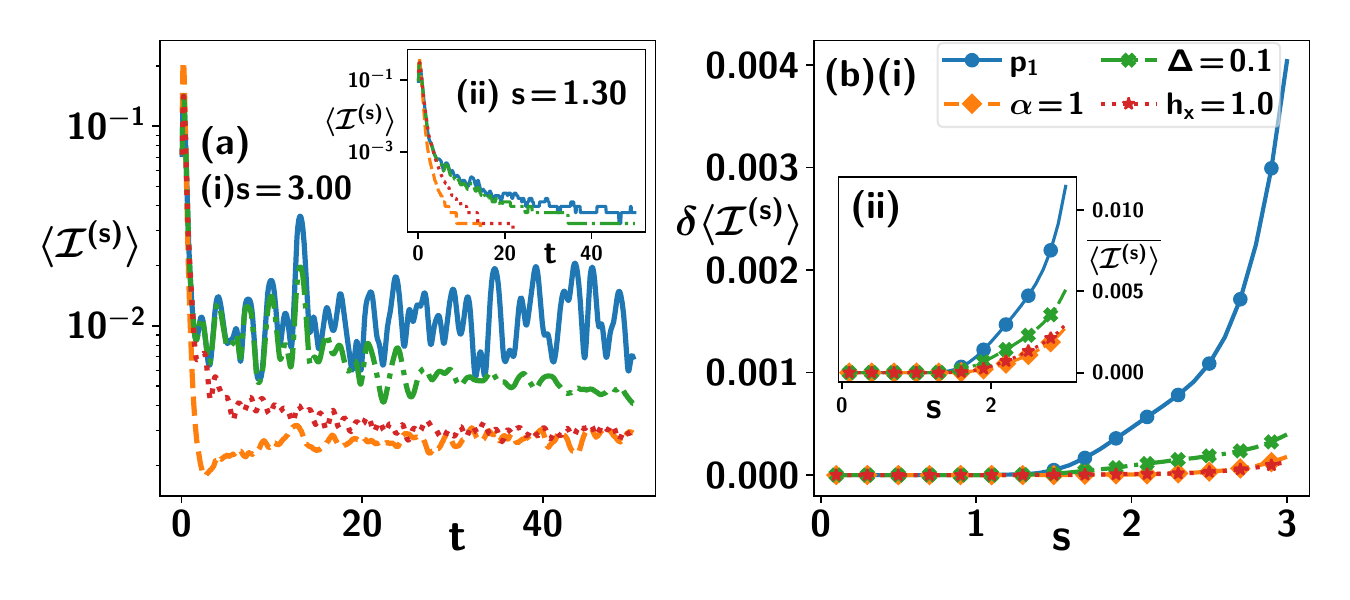}
   \caption{{\bf Efficacy of aSTC under non-Markovian dephasing noise}. (a) $\langle\mathcal{I}^{(s)}\rangle$ (ordinate in log-scale)) against time (abscissa) with Ohmicity parameter $s=3.0$ (non-Markovian regime) and \(s=1.3\) (Markovian domain) in the inset. Here \(N=8\). When the dynamics is non-Markovian, due to the backflow of information from bath, $\langle\mathcal{I}^{(s)}\rangle$ will grow for all systems, and for integrable systems ($p_1$),   its temporal fluctuation persists for a large time while it is suppressed in the Markovian regime as seen in Figs. \ref{fig:smi_deph} and \ref{fig:otoc_deph}. %(Inset) When $s=1.30$, $\langle\mathcal{I}^{(s)}\rangle$ (ordinate) vs time (abscissa) produces result like maximum Markovian noise due to which all temporal fluctuation of integrable system is suppressed and $\langle\mathcal{I}^{(s)}\rangle$ decays very fast. 
  (b) $\delta\langle\mathcal{I}^{(s)}\rangle$ (ordinate) against $s$ (abscissa) (with $(t_0, t_1)=(30, 50)$) clearly establishes the benefit of non-Markovian dynamics and the effectiveness of aSTC power. With increasing $s$, the evolution operator changes from extreme Markovian to non-Markovian and $\delta\langle\mathcal{I}^{(s)}\rangle$ also substantially increases for the integrable Hamiltonian, and a negligible increment is seen for the chaotic cases. (Inset) The mean $\overline{\langle\mathcal{I}^{(s)}\rangle}$ (ordinate) against $s$ (abscissa) also mimics the transition from Markovian to non-Markovian with increasing $s$ although $\delta\langle\mathcal{I}^{(s)}\rangle$ turns out to be more effective than the mean. All axes are dimensionless. }
  \label{fig:smi_nm_spec}
\end{figure}

\begin{figure}\includegraphics[width=\linewidth]{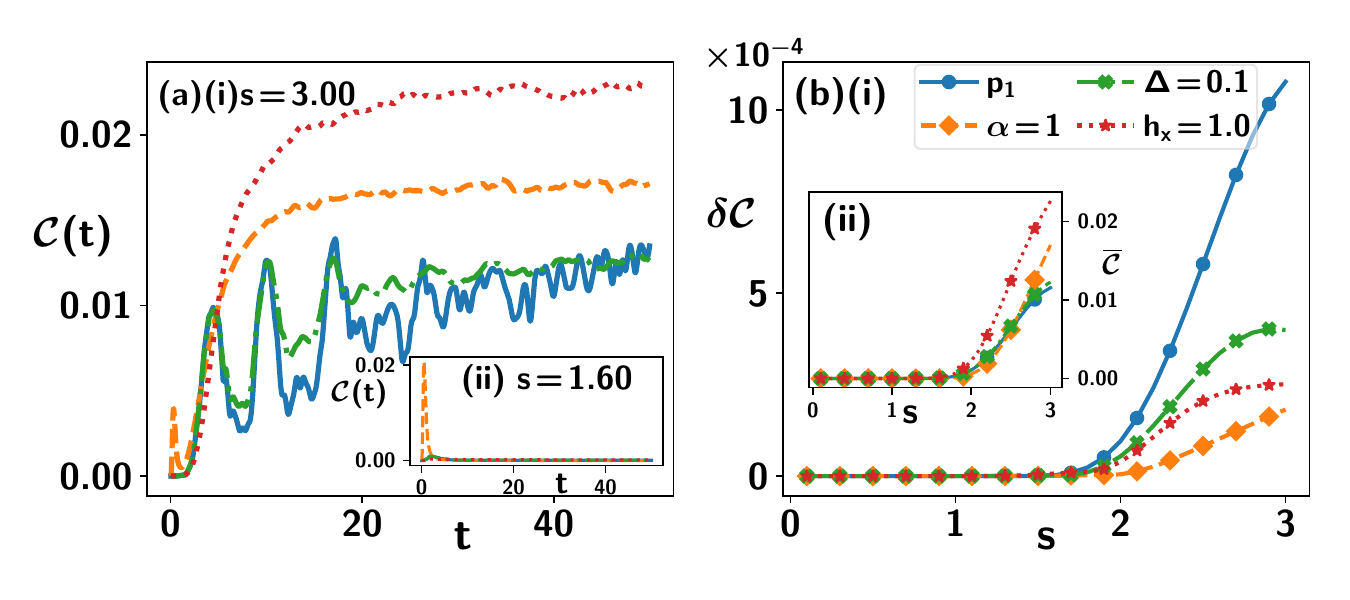}
  \caption{{\bf OTOC in the presence of spectral dephasing noise}. (a) $\mathcal{C}(t)$ (ordinate) against time (abscissa) for $s=3.0$. Chaotic systems ($\alpha=1$ and $h_x=1.0 $ ) produce fast-growing $\mathcal{C}(t)$ compared to the integrable one ($p_1$) and the chaotic one ($\Delta=0.1$). (Inset) The same plot for $s=1.6$ (Markovian). 
  %$\mathcal{C}(t)$ for all systems decays permanently, and temporal fluctuation vanishes.With $t_0=30$ and $t_1=50$, 
  (b) $\delta \mathcal{C}$ (ordinate) against $s$ (abscissa) (with $(t_0, t_1)=(30, 50)$) shows the transition of the system from Markovian to non-Markovian with increasing $s$. (Inset) The  mean $\overline{\mathcal{C}(t)}$ (ordinate) vs $s$(abscissa). Here \(N=8\) and all axes are dimensionless. }
  \label{fig:otoc_nm_spec}
\end{figure}

\begin{figure}\includegraphics[width=\linewidth]{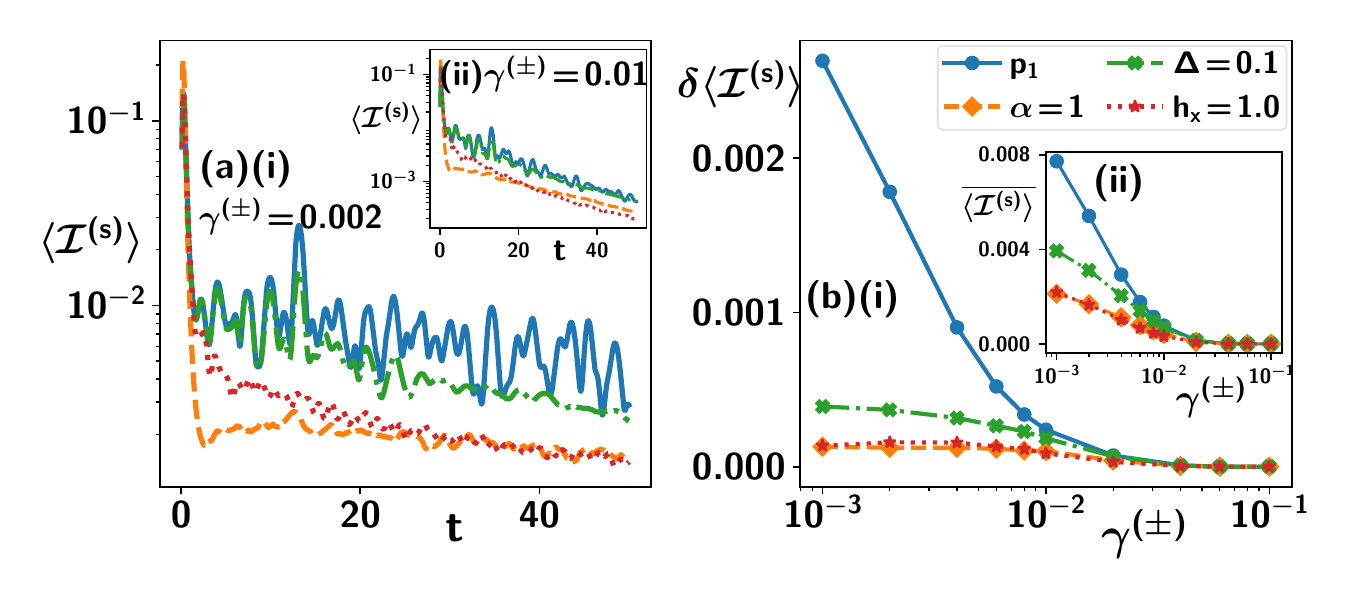}
  \caption{ {\bf Trade-off between non-Markovian  and Markovian noise.} With system-size $N=8$ and dephasing Ohmicity parameter, $s=3.0$, (a) $\langle\mathcal{I}^{(s)}\rangle$ (ordinate in log-scale) against time (abscissa) for a weak Markovian amplitude damping noise.  Due to the competition between Markovian and non-Markovian noise,  $\langle\mathcal{I}^{(s)}\rangle$ cannot decay fast, as seen in Fig. \ref{fig:smi_amp_damp}. Temporal fluctuation is present in the integrable system for a very long time. (Inset) When Markovian amplitude damping noise is high ($\gamma^{\pm}=0.01$), $\langle\mathcal{I}^{(s)}\rangle$ (ordinate) shows faster decay with time (abscissa) and faster suppression of temporal fluctuations in both chaotic or integrable systems is observed. (b)  Behavior of $\delta \langle\mathcal{I}^{(s)}\rangle$ (ordinate) with a amplitude damping decay rate $\gamma^{\pm}$ (abscissa) with \(s =3.0\) (and $(t_0, t_1)=(30, 50)$), highlighting the interplay between Markovian and non-Markovian noise. At weak Markovian noise ($\gamma^{\pm}\leq 10^{-2}$), for integrable systems, $\delta\langle\mathcal{I}^{(s)}\rangle \sim O(10^{-2})$ whereas for chaotic system, $\delta \langle\mathcal{I}^{(s)}\rangle \sim O(10^{-3})$. As $\gamma^{\pm}$ increases, $\delta \langle\mathcal{I}^{(s)}\rangle$ for both systems starts reducing and vanishing. (Inset) The mean $\overline{\langle\mathcal{I}^{(s)}\rangle}$ (ordinate) vs $\gamma^{\pm}$ (abscissa). 
  %In dominating non-Markovian noise $\overline{\langle\mathcal{I}^{(s)}\rangle}$ for the integrable system($p_1$) is large compared to all chaotic systems, the distinction between integrable and chaotic systems will vanish as Markovian noise (amplitude damping) starts to dominate. 
  All axes are dimensionless. }
  \label{fig:smi_nm_spec_s3}
\end{figure}

\begin{figure}\includegraphics[width=\linewidth]{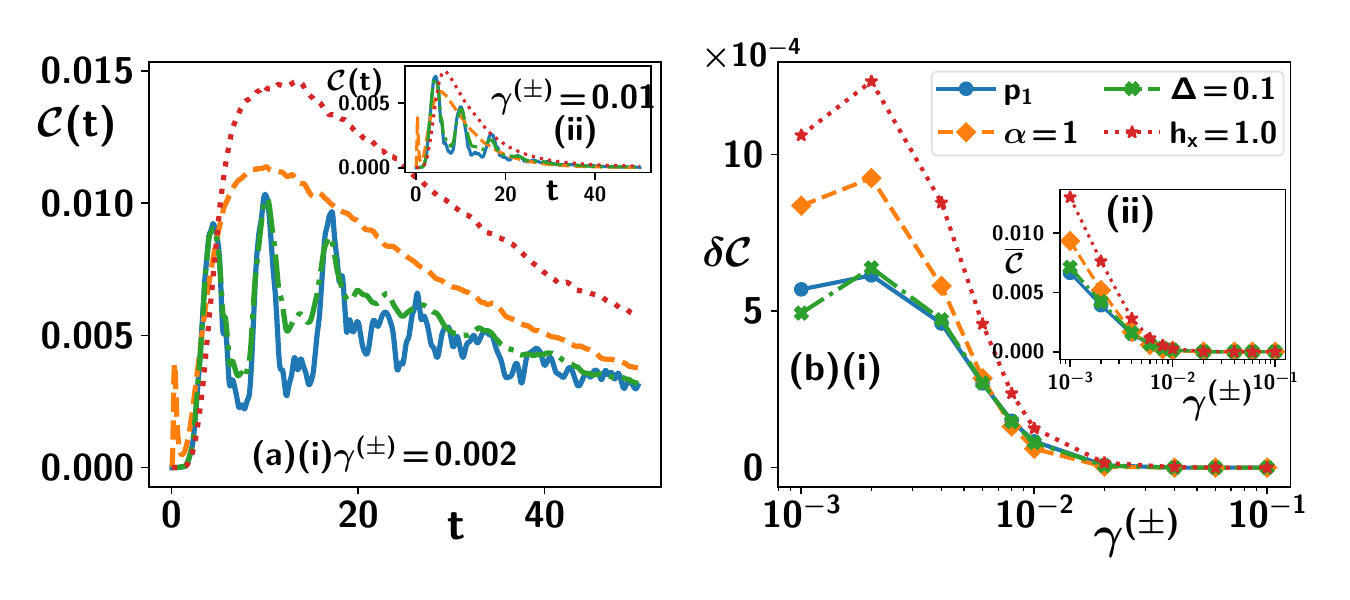}
  \caption{ {\bf Competition between Markovian and non-Markovian noise through OTOC}. The  considerations and parameters  are the same as in Fig. \ref{fig:smi_nm_spec_s3} except the physical quantity computed is OTOC.   
  %For the Ohmicity parameter $s=3.0$ and system-size $N=8$, (a) $\mathcal{C}(t)$ (ordinate) against time (abscissa) with weak Markovian amplitude damping noise $\gamma^{\pm}=0.002$. $\mathcal{C}(t)$ (for integrable or chaotic) grows initially due to the non-Markovian effect, but after a long time, due to the  Markovian noise, it again produces a decaying feature. Temporal fluctuation is present in both chaotic ($\Delta=0.1$) and integrable ($p_1$) cases.(Inset) At strong Markovian noise ($\gamma^{\pm}=0.01$) $\mathcal{C}(t)$ (ordinate) vs time (abscissa) plot denotes quick suppression of temporal fluctuation and a fast decaying feature. (b) Dependence of $\delta C$ on $\gamma^{\pm}$. At weak Markovian noise , for all systems(integrable and chaotic both) $\delta \mathcal{C}\sim O(10^{-4})$.(Inset) The mean $\overline{C}$ (ordinate) vs $\gamma^{\pm}$ (abscissa),$\overline{\mathcal{C}}\sim O(10^{-2})$ for all cases (integrable or chaotic). 
  All axes are dimensionless. }
  \label{fig:otoc_nm_spec_s3}
\end{figure}

\subsection{Time-dependent decay : Recovering the power of average STC}
\label{sec:ohmic_non_markov}

Let us first consider the time-dependent interactions between the system and the bath, specifically the dephasing bath, i.e.,  $L^{(z)}_j=\sigma^z_j$ at each site $j$, which arises from the system in contact
with a thermal bath of bosonic harmonic oscillators, having $a_\omega$ ($a_\omega^\dagger$) as the bosonic annihilation (creation) operator of the mode with energy $\omega$. The interaction between the thermal bosonic bath at the site $j$ is given by $H^{(j)}_{SE}=\sigma^z_j\otimes\int_\omega d\omega\left[ K(\omega) a_p + K(\omega)^* a_p^\dagger\right]$, with the spectral density
%\begin{equation}
\( K(\omega)=\frac{\omega^s}{\omega_c ^{s-1}}\exp(-\frac{\omega}{\omega_c})\),
%\end{equation}
where $\omega_c$ is the cut-off frequency of the reservoir, and $s$ is the Ohmicity parameter which describes the sub-Ohmic ($s<1$), Ohmic ($s=1$) and super-Ohmic ($s>1$) reservoirs~\cite{Haikka2013, Titas2018, Srijon2021}. With the bath initialized in the thermal state, the interactions results in the dynamics of the system described by the GKLS equations (Eqs.~(\ref{eq:gkls_st}) and~(\ref{eq:gkls_op})) with dephasing environment and the time-dependent system-bath couplings are given by 
\begin{equation}
\gamma^{(z)}(t,s)=[1+(\omega_c t)^2\bigr]^{-s/2} \Gamma(s) \sin\bigl[s\tan^{-1}(\omega_c t)\bigr],
\end{equation}
where $\Gamma(s)$ is the Euler Gamma function. The system describes a Markovian dynamics when $s\leq2$ and shows non-Markovian effects for $s>2$~\cite{Haikka2013, Titas2018, Srijon2021}.

The average STC power, $\langle\mathcal{I}^{(s)}\rangle$, shows a distinct behavior in the non-Markovian regime, i.e., $s>2$ for the integrable and chaotic Hamiltonians. In particular, in the non-Markovian domain, for the integrable systems with system-size $N=8$, the aSTC  oscillates around $\langle\mathcal{I}^{(s)}\rangle\sim10^{-2}$ in the long-time, while for the chaotic systems, $\langle\mathcal{I}^{(s)}\rangle\sim 10^{-3}$ with vanishing fluctuations, as shown in Fig.~\ref{fig:smi_nm_spec}. Hence in this case, the trend of average STC mimics the similar behavior illustrated with unitary dynamics. This is possibly due to the information backflow from the non-Markovian baths to the system. On the other hand, $\langle\mathcal{I}^{(s)}\rangle$ decays to zero in the Markovian dynamics with Ohmic baths $s\leq2$ as shown in the previous section. Note, further, that $\langle\mathcal{I}^{(s)}\rangle$ can distinguish the non-Markovian regime from the Markovian one. While the mean $\overline{\langle\mathcal{I}^{(s)}\rangle}$ shows discernible effects at Ohmicity $s$, it is system-size dependent (similar to case of unitary dynamics). Interestingly, the temporal fluctuations $\delta\langle\mathcal{I}^{(s)}\rangle$ of the aSTC power increase with the increasing non-Markovianity, and can faithfully distinguish the integrable and chaotic Hamiltonians in the non-Markovian situation.

The dynamics of the OTOC also captures the information reflux in the non-Markovian dynamics, with large temporal fluctuations in the dynamics with an integrable Hamiltonian, and a steady value for chaotic dynamics, at long times (see Fig.~\ref{fig:otoc_nm_spec}). Such a distinction is absent in the Markovian case, and the temporal fluctuations $\delta\mathcal{C}$ of the OTOC increase with increasing non-Markovianity. While $\overline{\mathcal{C}}$ also increases with increasing non-Markovianity, it fails to provide any distinction of integrable dynamics.

{\it Trade-off between Markovian and non-Markovian noise on integrability.  } To establish the role of non-Markovianity in the discrimination process of the Hamiltonians, along with time-dependent dephasing bath, we consider amplitude damping Markovian baths, i.e., $L^{(\pm)}_j=\sigma^{\pm}_j$ coupled to the system with strength $\gamma^{(\pm)}$ at each site $j$. This leads to a competition between Markovianity and non-Markovianity. In presence of moderate values of $\gamma^{(\pm)}$, the Markovianity dominates, leading to the eventual decay of aSTC at moderate times, and the information backflow from the non-Markovian environments cannot prevent the decay, as shown in Fig.~\ref{fig:smi_nm_spec_s3} for $s=3$. While the temporal fluctuations in average STC power in the integrable dynamics persist for small Markovian couplings $\gamma^{(\pm)}\lesssim 10^{-3}$ (with $t_0=30$ and $t_1=50$), it fails to distinguish between the integrable and chaotic dynamics at large $\gamma^{(\pm)}$, overwhelming the role of non-Markovian baths. Such adverse effects of Markovian baths are also present in the dynamics of OTOC. Surprisingly, the fluctuations of OTOC cannot properly distinguish integrable dynamics from the chaotic ones (see Fig.~\ref{fig:otoc_nm_spec_s3} for $s=3$ and $N=8$), which aSTC can faithfully do.

\begin{figure}\includegraphics[width=\linewidth]{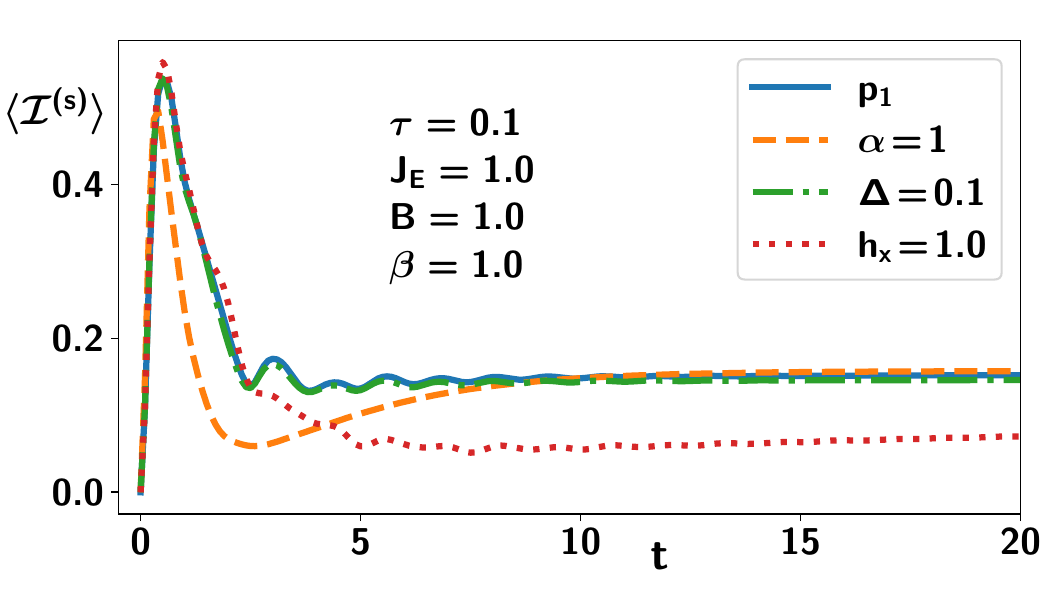}
  \caption{ {\bf Dynamics of aSTC in repetitive system-bath interaction. } $\langle\mathcal{I}^{(s)}\rangle$ (ordinate) against time (abscissa) with $\tau=0.1,J_E=1.0,B=1.0$ and $\beta=1.0$. Dynamics for the integrable system $p_1$ ((blue) solid line) and chaotic systems, namely $\Delta=0.1$ ((green) dash-dotted line),  $h_x=1.0$ ((red) dotted line), and \(\alpha =1\) ((orange) dashed lines). Trends of $\langle\mathcal{I}^{(s)}\rangle$  in the transient time can distinguish the integrable model with nearest-neighbor interactions from non-integrable ones with LR interactions.   All axes are dimensionless. }
  \label{fig:smirepit}
\end{figure}

\subsection{Strength of aSTC against repetitive system-bath interactions}

Let us now move to the investigation of finite-time interactions between the system and finite baths and its effect on the detection of integrability. Let us express the environment $E=\bigcup_{j=1}^{N}E^{(j)}$, where each local bath $E^{(j)}$ on site $j$, is an ensemble of $M^{(j)}$ qubits, i.e., $E^{(j)}=\{d^{(j)}_k,k=1, 2,\dots, M^{(j)}\}$. Each qubit $d^{(j)}_k$, with a local Hamiltonian $H_E^{(j,k)}$, is prepared in an initial state $\Omega_{d^{(j)}_k}$ and interacts with the site $j$ via Hamiltonian $H_{SE}^{(j,k)}$, for a finite time from $(k-1)\tau$ to $k\tau$ only~\cite{Titas2018}. Note that $M^{(j)}=0$ denotes the absence of the bath at site $j$, whereas $M^{(j)}=T/\tau$ for evolution upto time $T$. Therefore, the dynamics of the system are given by
\begin{align}
    \rho_{k\tau} &= \text{Tr}_E\left[ U_{\text{tot}}^{(k, \tau)} \left(\rho_{(k-1)\tau} \bigotimes_{j=1}^N \Omega_{d^{(j)}_k}\right) U_{\text{tot}}^{(k, \tau) \dagger}\right]\\
   U_{\text{tot}}^{(k, \tau)} &= \exp\left(-i H_{\text{tot}}^{(k)} \tau\right) \nonumber\\  H_{\text{tot}}^{(k)} &= H + \sum_j H_E^{(j,k)} + \sum_j H_{SE}^{(j,k)}, \nonumber
\end{align}
where $H$ describes the Hamiltonian of the system, as given in Eq.~(\ref{eq:H}). In this work, each bath qubit $d^{(j)}_k$ is taken to be a thermal state \( \Omega_{d^{(j)}_k} =\exp(-\beta H_E^{(j,k)})\big/\text{Tr}\left[\exp(-\beta H_E^{(j,k)})\right],\) of the local Hamiltonian $H_E^{(j,k)}$, 
\begin{align*}
    H_E^{(j,k)}=B\sigma^z_{d^{(j)}_k},
\end{align*}
where $B$ is the local magnetic field, and $\beta$ denotes the inverse temperature \(T\) of the bath.
The system-bath interactions are governed by the Hamiltonian,
\begin{align*}
    H_{SE}^{(j,k)} &= J_{SE}\left(\sigma^x_j\sigma^x_{d^{(j)}_k} + \sigma^y_j\sigma^y_{d^{(j)}_k}\right);
\end{align*}
where $J_{SE}$ is the interaction strength of the  bath with the system. Note that in this scenario, the system follows the amplitude damping Markovian dynamics (given in Eqs.~(\ref{eq:gkls_st}) and in Appendix  ref{}), when $\tau\to0$~\cite{Titas2018}. Such a decoherence model is referred to as repetitive  or the collisional models. 

The dynamics of the system in the presence of repetitive interactions show decreased temporal fluctuations due to chaotic dynamics of the system-bath Hamiltonian $H^{(k)}_{\text{tot}}$ for each step $k$. We consider repetitive baths on half of the system, i.e., $M^{(j)}=0$ for $j=1,2,\ldots, N/2-1$, and the rest are attached to the thermal baths. 
%dynamics of the  aSTC  are illustrated in Fig.~\ref{fig:smi_lrqi} for environmental parameters $J_E=\beta=B=1.0$ and finite-time $\tau=0.1$ and system-size $N=6$. 
We observe that in this collisional model, the detection of integrability is not as pronounced as can be seen for the dephasing noise. Specifically, the fluctuations of the integrable systems with parameter set $p_1$ persist for a small time $t\sim 5$, along with the chaotic systems with $\Delta=0.1$, and with $h_x=1.0$ (see Fig. \ref{fig:smirepit}). However, for long-range interactions, the aSTC possesses vanishing fluctuations, and it saturates to a parameter-dependent value. Such a behavior is observed for varying environmental parameters ($J_E, B$ and $\beta$) and bath interaction time $\tau$. Therefore, in the present system--reservoir setting, the aSTC exhibits a weaker discriminating capability than in dephasing and amplitude damping baths.  This behavior is likely due to thermal noise, which progressively suppresses the signatures imprinted by the evolution operator.

%Therefore, detecting integrability is unfeasible in general non-Markovian dynamics.  

% The dynamics is Markovian when $\tau\ll 1$ and show non-Markovian effects for $\tau\gtrsim1$~\cite{sen2023noisybattery}. 

\subsection{Initial mixed state creation via non-Markovian dynamics  }

Upto now, we always prepared the initial  states as Haar randomly chosen pure states and then the system evolved according to the Hamiltonian. Let us consider the situation when the preparation process is influenced by noise, i.e., the initial state can be given by a mixed state.
When the non-Markovian dephasing noise affects the initial state preparation, i.e., the Kraus operators affecting the states are given by~\cite{Shrikant_2018, Muhuri2024}
\begin{equation*}
K_I^{\mathrm{dph}} = \sqrt{(1-\chi q)(1-q)}\,\mathbb{I}, \quad
K_z^{\mathrm{dph}} = \sqrt{[1+\chi(1-q)]q}\,\sigma_z,
% \label{eq:dph}
\end{equation*}
while for the non-Markovian depolarizing channel, the Kraus operators take the form
\begin{equation*}
K_I^{\mathrm{dp}} = \sqrt{(1-3\chi q)(1-q)}\,\mathbb{I}, \quad
K_\mu^{\mathrm{dp}} = \sqrt{\frac{[1+3\chi(1-q)]q}{3}}\,\sigma^\mu,
% \label{eq:dp}
\end{equation*}
where $\mu \in \{x,y,z\}$ and $\chi$ is the non-Markovian parameter. To keep the complete positivity condition, for dephasing $q\in[0, 0.5]$, while for depolarizing, $0 \leq q \leq \frac{1}{3\chi}$, with $\chi=0$, giving the Markovian limit. Finally, the noisy initial states evolve with integrable and non-integrable Hamiltonians for a single realization, and we compute again the Haar averaged sum of total correlations. Due to Haar averaging, the discriminating capability of aSTC  reported before does not alter with the initial mixed states.

%Such a noise in the preparation of the mixed state does not change the previous analysis, due to the Haar-average over the initial pure states.

\section{Conclusion}
\label{sec:conclusion}

Determining whether a quantum many-body system is integrable or nonintegrable is a fundamental problem with  implications in both quantum physics and quantum technologies. A powerful framework for probing this distinction is provided by quantum information scrambling, which characterizes the process by which initially localized quantum information, encoded in a fully separable state, becomes distributed into highly nonlocal  correlations under the action of an interacting Hamiltonian. It was shown that integrable and nonintegrable systems exhibit qualitatively different scrambling dynamics --  chaotic systems rapidly delocalize quantum information while integrable systems display slower, and more structured spreading of information. 

Our work proposes quantum information theoretic quantities of quantum information scrambling based on the sum of total correlations (STC), which captures both classical and quantum correlations, together with genuine multipartite entanglement (GME) measures generated dynamically from initially fully separable states. Starting from Haar-random product states, we investigated the unitary dynamics of the Haar-averaged STC (aSTC) and GME in both integrable and nonintegrable quantum spin models, namely variable-range quantum \(XYZ\)  spin models in the presence of longitudinal and transverse magnetic fields. We demonstrated that the long-time dynamics of the aSTC and GME exhibit qualitatively distinct behavior in the two regimes. In particular, integrable systems display pronounced long-time temporal fluctuations, whereas chaotic systems rapidly approach a steady behavior with strongly suppressed fluctuations. Although the long-time mean of aSTC itself also distinguishes the two dynamical regimes, we showed that its temporal fluctuations provide a system-size-independent indicator of integrability, thereby serving as an information-theoretic identifier of integrability analogous to the OTOC.

Another important direction addressed in this work is whether the information theoretic quantities remain effective even when the system interacts with the environment, a regime that has remained largely unexplored in this context. We demonstrated that the discriminating capability of the aSTC persists in the presence of noise, thereby revealing the role of environmental effects in the scrambling of quantum information. In the presence of Markovian baths, information is continuously lost to the environment, leading to a progressive decay of the aSTC at long times. Nevertheless, we showed that for weak system-bath couplings, the temporal fluctuations of the aSTC in the transient time continue to distinguish integrable from chaotic dynamics. In contrast, under non-Markovian dynamics, the backflow of information from the environment revives the scrambling process, enabling both the aSTC and the OTOC to recover their ability to distinguish integrable and chaotic dynamics. This demonstrates that environmental memory effects can restore the signatures of quantum information scrambling through information backflow.
To investigate the impact of environmental memory on quantum information scrambling, we analyzed the dynamics under Markovian amplitude damping and non-Markovian dephasing channels. We found that the temporal fluctuations of the aSTC remain a robust discriminator of integrable and non-integrable dynamics in the weak Markovian regime, where the OTOC loses its distinguishing capability. Our results therefore establish the temporal fluctuations of the aSTC as a robust information-theoretic probe of quantum information scrambling in open quantum systems.

Beyond identifying faithful information-theoretic quantifiers capable of distinguishing integrable and chaotic dynamics, our work exhibits that these signatures remain remarkably robust in the presence of environmental interactions. In particular, the persistence of their discriminating capability under non-Markovian dynamics, which provides a more realistic description of many experimental platforms, highlights the resilience of information scrambling against decoherence when environmental memory effects are present. We, therefore, can expect our results to provide a useful framework for investigating information scrambling, quantum chaos, and integrability in realistic quantum simulators and processors, while offering new perspectives for the design of noise-resilient quantum information processing protocols.

\acknowledgements

 We acknowledge the use of the cluster computing facility at the Harish-Chandra Research Institute. This research was supported in part by the INFOSYS scholarship for senior students. We acknowledge support from the project entitled "Technology Vertical - Quantum Communication'' under the National Quantum Mission of the Department of Science and Technology (DST)  (Sanction order No. Sanction Order No. DST/QTC/NQM/QComm/2024/2 (G)). 

\appendix

% \section{Multipartite entanglement calculation}
% \label{app:gm}

% As mentioned in the main text, to mitigate the computational  challenge in GGM, we simplify our approach by restricting attention to Schmidt coefficients obtained from single-site reduced density matrices. With this simplification, we define the geometric measure (GM), also a multipartite entanglement measure, as follows:
% \begin{align}
%     \mathcal{G}_1(| \Psi\rangle)=1-\max_{A:B,|A|=1}[\lambda_{A:B}^2|A\cap B=\emptyset,|A\cup B|=N],
% \end{align}
% where the maximization is performed only over bipartitions for which \(A\) consists of a single spin. To validate the effectiveness of this approximation, we compute both \(\mathcal{G}\) and \(\mathcal{G}_1\) for various system-sizes \(N\), averaging over the middle of the Hamiltonian spectrum. As shown in Fig. \ref{fig:ggm_vs_gm}, the average difference \(|\overline{\langle \mathcal{G}\rangle} - \overline{\langle \mathcal{G}_1\rangle}|\approx e^{-N}\) is found to be very small and it decreases rapidly with increasing system-size, indicating that the \(\mathcal{G}_1\) measure is as effective as \(\mathcal{G}\) in characterizing multipartite entanglement. 
%Therefore, in this paper, we use \(\mathcal{G}\) and \(\mathcal{G}_1\) alternatively as there is no such distinction between these two quantities to quantify MBL phases which can help in studying the role of two- and three-body DM interaction in altering the phases in Heisenberg spin chain, which we study in the next section.

\section{Time evolution with Chebyshev method}
\label{app:chb}

The Chebyshev polynomial expansion provides an efficient approach for computing the time evolution of quantum states, particularly for systems with large Hilbert spaces where exact diagonalization is computationally expensive~\cite{Weise2006, Weise2008}. For a Hamiltonian $H$ and an initial state $|\psi(0)\rangle$, the state at time $t$ is written as
\[
|\psi(t)\rangle = e^{-iHt}|\psi(0)\rangle .
\]

Instead of evaluating the exponential operator directly, the evolution operator can be expressed as a series of Chebyshev polynomials of the first kind, which are defined on the interval $[-1,1]$. To apply this expansion, the Hamiltonian is first rescaled so that its eigenvalues lie within this interval, with \(\tilde{H} = \frac{H-b}{a}\), where, $a=(E_{\max}-E_{\min})/2$ and $b=(E_{\max}+E_{\min})/2$, with $E_{\min}$ and $E_{\max}$ denoting the smallest and largest eigenvalues of $H$ and can be computed using Lanczos algorithm. The time-evoluted state can be approximated as
\[
\ket{\psi(t)} \approx e^{-ibt}\left(
J_0(at) + 2\sum_{n=1}^{K} (-i)^n J_n(at)\, T_n(\tilde{H})
\right)\ket{\psi_0}
\]
where $J_n(t)$ is the Bessel function of order $n$ , $T_n(x)$ is the Chebyshev polynomial of order $n$ and $\ket{\psi_0}$ is the initial state~\cite{Sierant2019}. The approximation is exact as $K\to\infty$, and a finite $K$ is chosen such that the $K$-th term has magnitude $<10^{-8}$. The Chebyshev polynomials satisfy the recurrence relation
\[
T_n(x) = 2x\,T_{n-1}(x) - T_{n-2}(x),
\]
with $T_0(x)=1$ and $T_1(x)=x$. Let us define \(\ket{v_n(t)} = T_n(\tilde{H}) \ket{\psi_{0}}\), then the recurrence relation for the Chebyshev polynomial to compute the time evolved state is given by
\[
\ket{v_n(t)} = 2\tilde{H}\,\ket{v_{n-1}(t)} - \ket{v_{n-2}(t)},
\]
with \(\ket{v_0(t)} = \ket{\psi_{0}}\), \(\ket{v_1(t)} = \tilde{H}\ket{\psi_{0}}\).

For long-time evolution, a very large value of $N$ is required, which is computationally difficult. Therefore, we evolve the state using a small time interval $dt=0.1$ and $K=30$ provides an approximately accurate unitary evolution in our case.

\section{Average genuine multipartite entangling power for detection of integrability in noiseless scenario}
\label{app:ggm}

\begin{figure}[h]
  \includegraphics[width=1\linewidth]{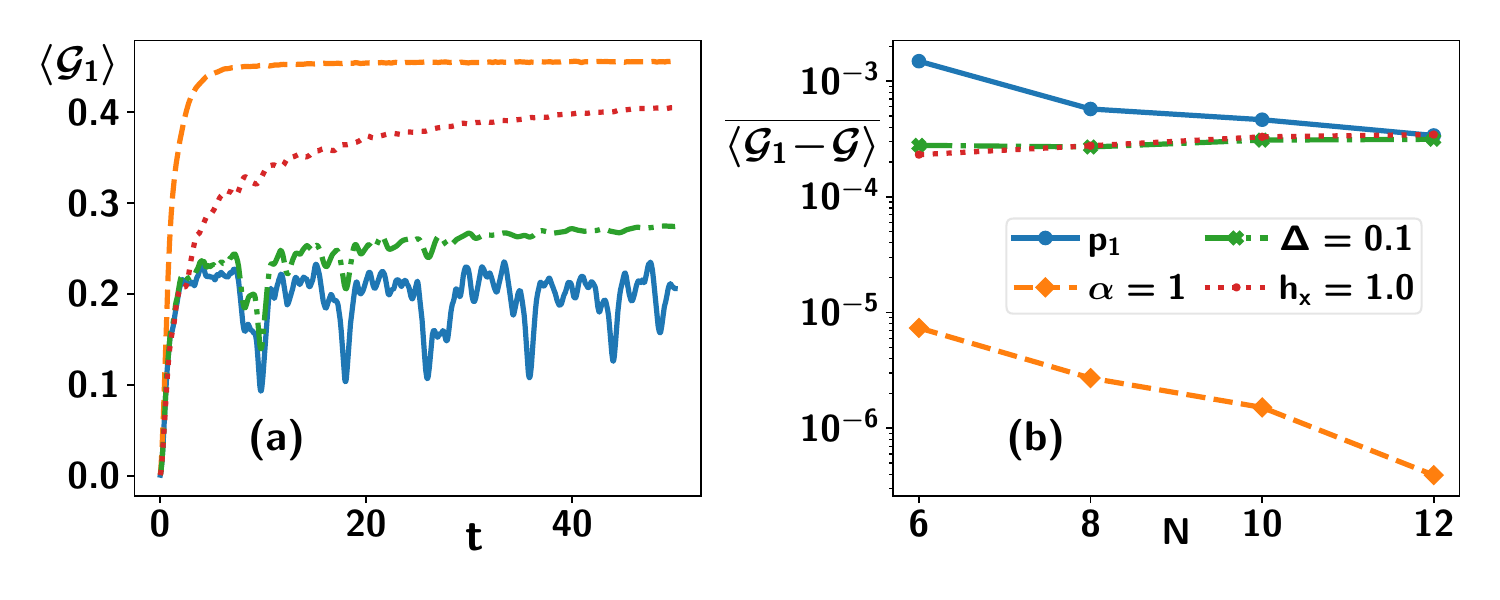}
  \caption{{\bf Dynamics of GGM for unitary dynamics}. For the system-size $N=18$ and ensemble of $100$ initial pure product states, (a) $\langle \mathcal{G}_{1}\rangle$ (ordinate) against time (abscissa). Temporal fluctuation remains in the integrable system $p_1$ ((blue) solid line) at very long times, where all chaotic systems produce very fast-growing and saturating  $\langle \mathcal{G}_{1}\rangle$. (b) The average difference $\overline{\langle \mathcal{G}_{1}-\mathcal{G}\rangle}$ (ordinate in log-scale) vs system-size \(N\) (abscissa) shows that for large system-size or chaotic cases, we do not need to consider all possible bipartitions to compute GGM; only a single-site contribution is enough. 
  %(c) Dependence  $\delta \langle \mathcal{G}_{1}\rangle $ (ordinate) on $h_x$ (abscissa) and $\alpha$(abscissa) (with $t_0=30$ and $t_1=50$).(Inset) The mean $\overline{\langle \mathcal{G}_{1}\rangle}$ (ordinate) vs $h_x$ (abscissa) and $\alpha$ (abscissa). 
  All axes are dimensionless.}
  \label{fig:ggm_uni}
\end{figure}

\begin{figure}[h]
  \includegraphics[width=\linewidth]{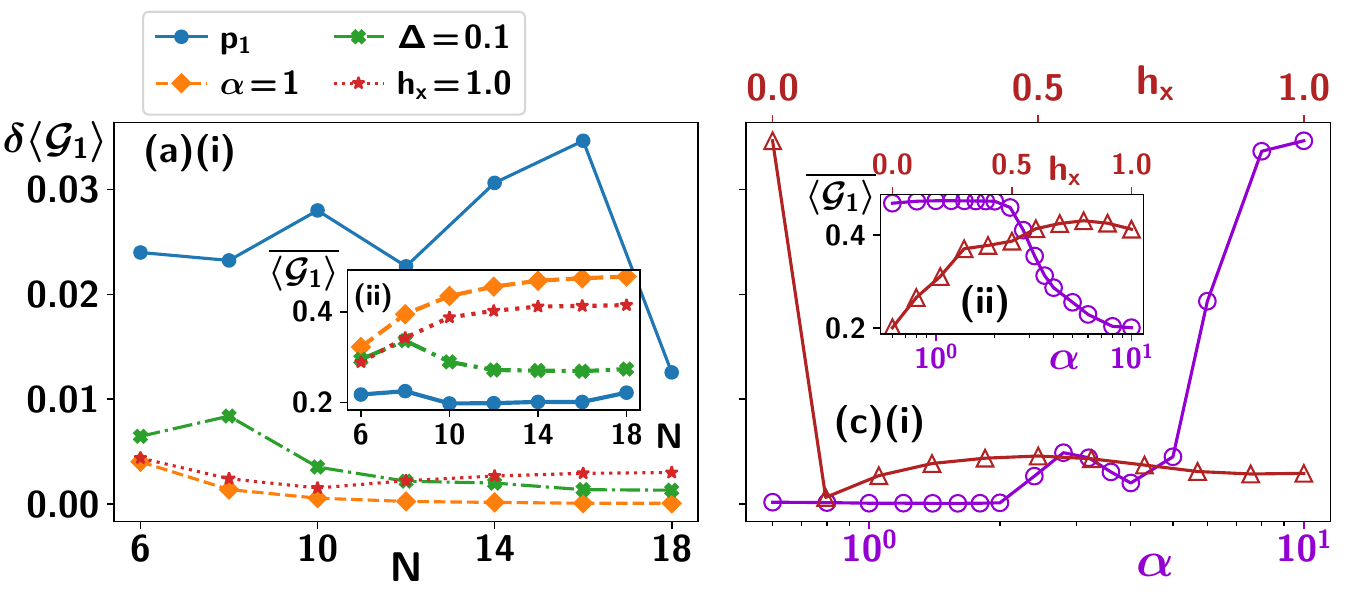}
  \caption{ {\bf System-size dependent behavior of GGM.} (a) $\delta\langle \mathcal{G}_{1}\rangle$ (ordinate)  vs N (abscissa). (b) $\delta{\langle \mathcal{G}_{1}\rangle}$ (ordinate) vs \(\alpha\) (lower abscissa) and \(h_x\) (upper abscissa). (Inset) (a)(ii) and (b)(ii) $\overline{\langle \mathcal{G}_{1}\rangle}$ (ordinate) with respect to the same horizontal axis in (a)(i) and (b)(i). All axes are dimensionless.}
  \label{fig:ggm_uni_N}
\end{figure}

In the noiseless scenario, i.e., $\gamma^{(k)}_j=0 \ \forall j$ in Eqs.~(\ref{eq:gkls_st}) and~(\ref{eq:gkls_op}) of the main text, the system is isolated, and the dynamics is described by a unitary evolution. In this section, we study the average genuine multipartite entangling power of the Hamiltonian $H$ in Eq.~(\ref{eq:H}) of the main text to distinguish the integrable system from the chaotic one. Within the framework studied in the work, the initial states of the system are taken as pure product states chosen independently and Haar randomly over qubit state space at each site. Therefore, the dynamical states in unitary evolution are also pure states, and the genuine multipartite entanglement (GME) can be quantified by generalized geometric measure (GGM)~\cite{GGM_wei2003, aditi_pra_2010, Samanta_2026}. A pure state of $N$ parties is genuinely entangled, if it is not separable in any bipartitions of the system. GGM is a distance based measure of GME, defined as the minimum distance from the set of all bi-separable states. Mathematically, the GGM $\mathcal{G}(\ket{\Psi})$ of a pure state $\ket{\Psi}$ is computed as 
\begin{align}
   \mathcal{G}(\ket{\Psi}) &= 1-\underset{A, B, \ket{\phi_A}, \ket{\phi_B}}{\max}\bigg|\big(\bra{\phi_A}\!\otimes\!\bra{\phi_B}\big)\ket{\Psi}\bigg|^2 \nonumber \\
   &= 1-\max_{A:B} [\lambda_{A:B}^2],
\end{align}
where \(\lambda_{A:B}\) denotes the largest Schmidt coefficient of a given bipartition \(A:B\), and the maximization is carried out over all bipartitions \(A:B\) such that $A\cap B=\emptyset$ and $|A\cup B|=N$. As the number of bipartitions increases exponentially with increasing system-size, we approximate the GGM $\mathcal{G}(\ket{\Psi})$ by $\mathcal{G}_1(\ket{\Psi})$, computed by maximization over bipartitions for which \(A\) which consists of a single site only, i.e., $|A|=1$, with $\mathcal{G}_{1}(t) \equiv \mathcal{G}_{1}(\ket{\Psi(t)})$.

The average multipartite entangling power, denoted by $\langle\mathcal{G}_1\rangle(t)$, is the average GGM generated in the initially Haar random product state, via Hamiltonian $H$. We observe that the $\langle\mathcal{G}_1\rangle(t)$ increases with time initially, and at long time, saturates to a parameter dependent value for chaotic Hamiltonian, whereas it oscillates around a mean value for integrable systems, as shown in Fig.~\ref{fig:ggm_uni}(a) for parameters $p_1\equiv\{\gamma=0.5, \alpha=10, h_z=1, \Delta=h_x=0\}$ and parameters different from $p_1$, specifically $\alpha=1$, $\Delta=1$ or $h_x=1$. To validate the approximation of GGM by $\mathcal{G}_1$, we compute the difference $\mathcal{G}_1(t)-\mathcal{G}(t)$ for the dynamical states . We observe that the Haar average of difference $\langle\mathcal{G}_1(t)-\mathcal{G}(t)\rangle\lesssim10^{-3}$ for system-size $N\geq 6$, and the temporal mean $\overline{\langle\mathcal{G}_1-\mathcal{G}\rangle}$ of the difference decreases exponentially with the system-size $N$ (see Fig.~\ref{fig:ggm_uni}(b)).

The fluctuating behavior of GGM for integrable dynamics gives the quantifier, $\delta\langle\mathcal{G}_1\rangle$ (defined from the second moment in Eq.~(\ref{eq:moments}) of the main text) for detecting the integrable and chaotic dynamics. For integrable dynamics the temporal fluctuations $\delta\langle\mathcal{G}_1\rangle\sim 0.02$, remain independent of the system-size. On the contrary, for chaotic systems, $\delta\langle\mathcal{G}_1\rangle\lesssim 10^{-3}$, thereby distinguishing the integrable dynamics from chaotic ones (see Fig.~\ref{fig:ggm_uni_N}(a)). For integrable systems, non-integrability is introduced with non-vanishing longitudinal fields $h_x\neq0$ and long-range interactions $\alpha\sim 1$. The temporal fluctuations captures the non-integrable dynamics with $\delta\langle\mathcal{G}_1\rangle\lesssim 10^{-3}$ for $h_x>0$ and $\alpha\lesssim2$, and increasing for increasing $\alpha$ giving integrable limit (see Fig.~\ref{fig:ggm_uni_N}(b)). The temporal mean $\overline{\langle\mathcal{G}_1\rangle}$ is both parameter and system-size dependent, and fails to discern the integrable and chaotic dynamics.

\section{Information scrambling in quantum spin chain with amplitude damping baths}
\label{app:amp_damp}

\begin{figure}
  \centering
  \includegraphics[width=\linewidth]{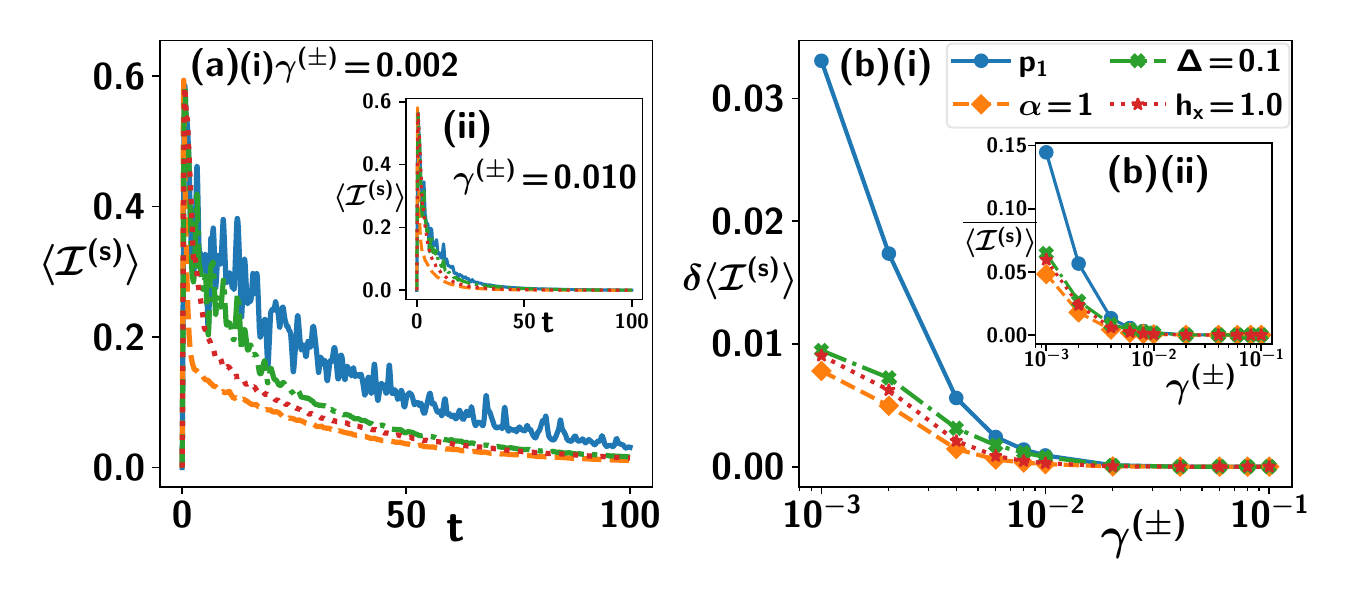}
  \caption{ {\bf Dynamics of average STC in the presence of Markovian amplitude damping noise.} For the system-size $N=8$ (a) $\langle\mathcal{I}^{(s)}\rangle$ (ordinate) against time (abscissa) for weak Markovian amplitude damping noise ($\gamma^{\pm}=0.002$),behavior is very analogous to Fig.~\ref{fig:smi_deph} (a). (Inset) $\langle\mathcal{I}^{(s)}\rangle$ (ordinate) vs time (abscissa)  for $\gamma^{\pm}=0.01$. (b) Dependence of $\delta \langle\mathcal{I}^{(s)}\rangle$ (ordinate) on $\gamma^{\pm}$ (abscissa).(Inset) The mean $\overline{\langle\mathcal{I}^{(s)}\rangle}$ (ordinate) vs $\gamma^{\pm}$ (abscissa). All axes are dimensionless.}
  \label{fig:smi_amp_damp}
\end{figure}

\begin{figure}
  \includegraphics[width=\linewidth]{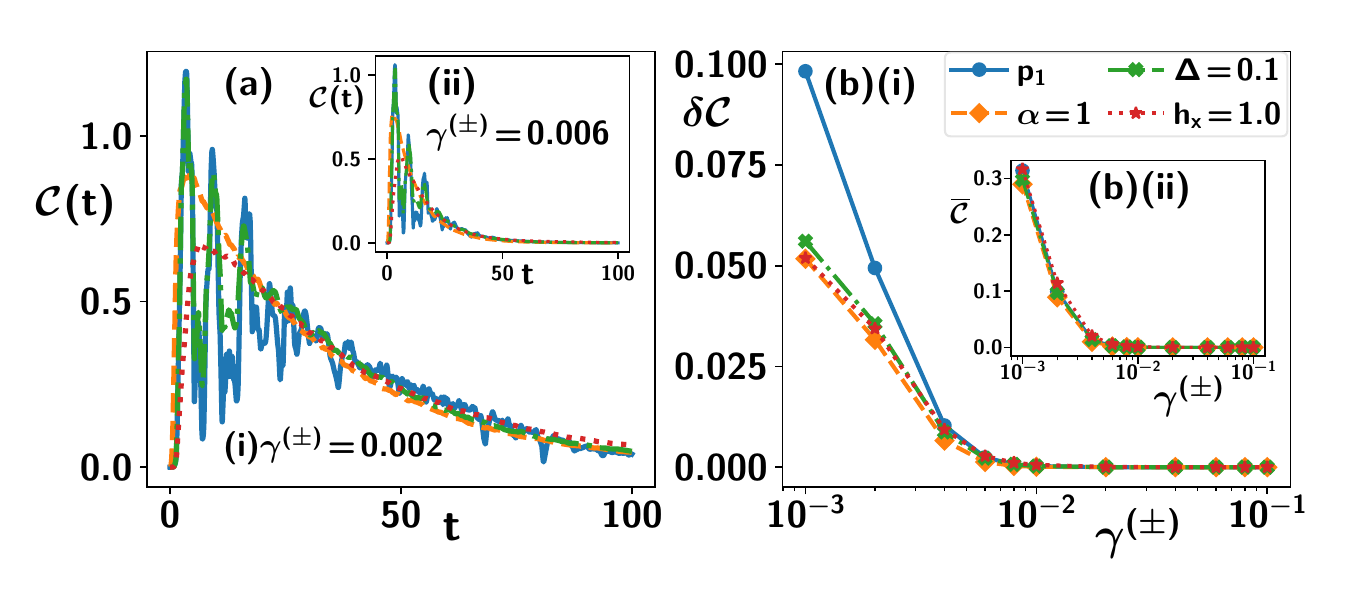}
  \caption{{\bf Dynamics of OTOC in the presence of amplitude damping noise.}(a) $\mathcal{C}(t)$ (ordinate) against time(abscissa) with $\gamma^{\pm}=0.002$ . (Inset) At strong amplitude damping noise ($\gamma^{\pm}=0.01$), $\mathcal{C}(t)$(ordinate) vs time (abscissa). With $t_0=60$ and $t_1=100$ (b) Dependence of $\delta \mathcal{C}$ (ordinate) on $\gamma^{\pm}$ (abscissa). (Inset) The mean $\overline{\mathcal{C}}$ (ordinate) vs time (abscissa). All axes are dimensionless. }
  \label{fig:otoc_amp_damp}
\end{figure}

In  Sec.~\ref{subsec:Markoviandph}, we discuss the distinction between the integrable and chaotic dynamics of the quantum system in the presence of Markovian dephasing noise. Here, we show that similar results emerge with the amplitude damping Markovian noise. In this situation, starting from the initial Haar uniformly generated product state, the system evolves according to the GKLS master equation in Eq.~(\ref{eq:gkls_st}) with the Lindblad operators $L^{(\pm)}_{j}=\sigma_{j}^{\pm}$, with (uniform) dissipation coefficient $\gamma^{(\pm)}$ and $\sigma^{\pm}_j = (\sigma^{x} \pm i\sigma^{y})/2$ at each site $j$.

In the presence of amplitude damping baths, the average STC power $\langle\mathcal{I}^{(s)}\rangle$ first increases due to the correlations generated by the interacting Hamiltonian. At long times, $\langle\mathcal{I}^{(s)}\rangle$ decreases with time, due to the loss of information to the Markovian baths (see Fig.~\ref{fig:smi_amp_damp}(a)). The fluctuations for the integrable system persist for a small amplitude damping noise strength $\gamma^{(\pm)}$. As $\gamma^{(\pm)}$ increases, $\langle\mathcal{I}^{(s)}\rangle$ decays faster with the suppression of the temporal fluctuations $\delta \langle\mathcal{I}^{(s)}\rangle$. Therefore, the temporal mean $\overline {\langle\mathcal{I}^{(s)}\rangle}$ and fluctuations $\delta \langle\mathcal{I}^{(s)}\rangle$ can distinguish the integrability of the system only at weak noise strength. At $\gamma^{\pm}\leq10^{-2}$, for the integrable model, $\delta \langle\mathcal{I}^{(s)}\rangle \sim O(10^{-2})$ and $\overline{\langle\mathcal{I}^{(s)}\rangle}\sim O(10^{-2})$, while for  chaotic models $\delta \langle\mathcal{I}^{(s)}\rangle \sim O(10^{-3})$ and $\overline{\langle\mathcal{I}^{(s)}\rangle}\sim O(10^{-3})$.

The operator dynamics given by the GKLS equation, Eq.~(\ref{eq:gkls_op}) provides the behavior of OTOC in the presence of amplitude damping noise. Similar to the dephasing noise, the fluctuations in OTOC for the integrable system decrease. With increasing $\gamma^{\pm}$, $\mathcal{C}(t)$ decays faster with time, and the temporal fluctuations are also reduced. At $\gamma^{\pm}\leq 10^{-2}$, for the integrable case, $\delta \mathcal{C}\sim O(10^{-2})$ while for the chaotic models, $\delta \mathcal{C}\sim O(10^{-2})$ and on the other hand, in both systems (chaotic and integrable) $\overline{\mathcal{C}(t)}\sim O(10^{-2})$ (see Fig.~\ref{fig:otoc_amp_damp}).

\bibliography{ref.bib}
\end{document}